\documentclass[twocolumn]{aa}
\usepackage{graphicx}
\def\hi{H{\sc i\,}\ }

\def\kms{km~s$^{-1}$}

\def\farcm{\hbox{$.\mkern-4mu^\prime$}}
\usepackage{txfonts}
\usepackage{natbib}
\bibpunct{(}{)}{;}{a}{}{,}
\newcommand{\msun}{$\rm M_{\odot}$}
\newcommand{\Lsun}{$\rm L_{\odot}$}
\newcommand{\mhi}{$ M_{HI}$}
\newcommand{\lb}{$\rm L_{B}$}

\begin{document}
 \title{Stars and gas in the Medusa merger}

   \subtitle{}

   \author{E. Manthey\inst{1,2}
	  \and
	  S. H\"uttemeister\inst{2}
          \and
          S. Aalto\inst{3} 
	  \and
	  C. Horellou\inst{3}
	  \and
          P. Bjerkeli\inst{3}
          }

   \offprints{E. Manthey}

   \institute{
	Astron, 7990AA Dwingeloo, Netherlands\\
	 \email{juette@astron.nl}
	\and
	University of Bochum, Department of Astronomy,
            44780 Bochum, Germany\\ 
         \and
             Onsala Space Observatory, Chalmers University of Technology, SE-43992 Onsala, Sweden
             }

   \date{Received; accepted}

   \abstract{
The Medusa (NGC~4194) is 
a well-studied nearby galaxy with the disturbed appearance
of a merger and evidence for ongoing star formation.
In order to test whether it could be the result of an interaction between a gas-rich disk-like galaxy 
and a larger elliptical, we have carried out optical and radio observations of the stars and 
the gas in the Medusa, and performed $N$-body numerical simulations of the evolution of such a system. 
We used the Nordic Optical Telescope to obtain a deep V-band image 
and the Westerbork Radio Synthesis Telescope to map the large-scale distribution
and kinematics of atomic hydrogen.
A single \hi tail was found to the South of the Medusa with a projected length
of $\sim 56$~kpc ($\sim 5'$) and a gas mass of $7\cdot 10^8\,M_{\odot}$, 
thus harbouring about one third of the total \hi mass of the system. 
\hi was also detected in absorption toward the continuum in the center. 
\hi was detected in a small nearby galaxy to the North-West of the Medusa 
at a projected distance of 91\,kpc. 
It is, however, unlikely that this galaxy
has had a significant influence on the evolution of the Medusa.
The simulations of the slightly prograde infall of a gas-rich disk galaxy on an larger, 
four time more massive elliptical (spherical) galaxy reproduce most of the observed 
features of the Medusa.Thus, the Medusa is an ideal object to study the merger-induced
star formation contribution from the small galaxy of a minor merger.

  } 

   \maketitle
%________________________________________________________________

\section{Introduction}

\begin{figure*}
   \includegraphics[angle=270,width=8.5cm]{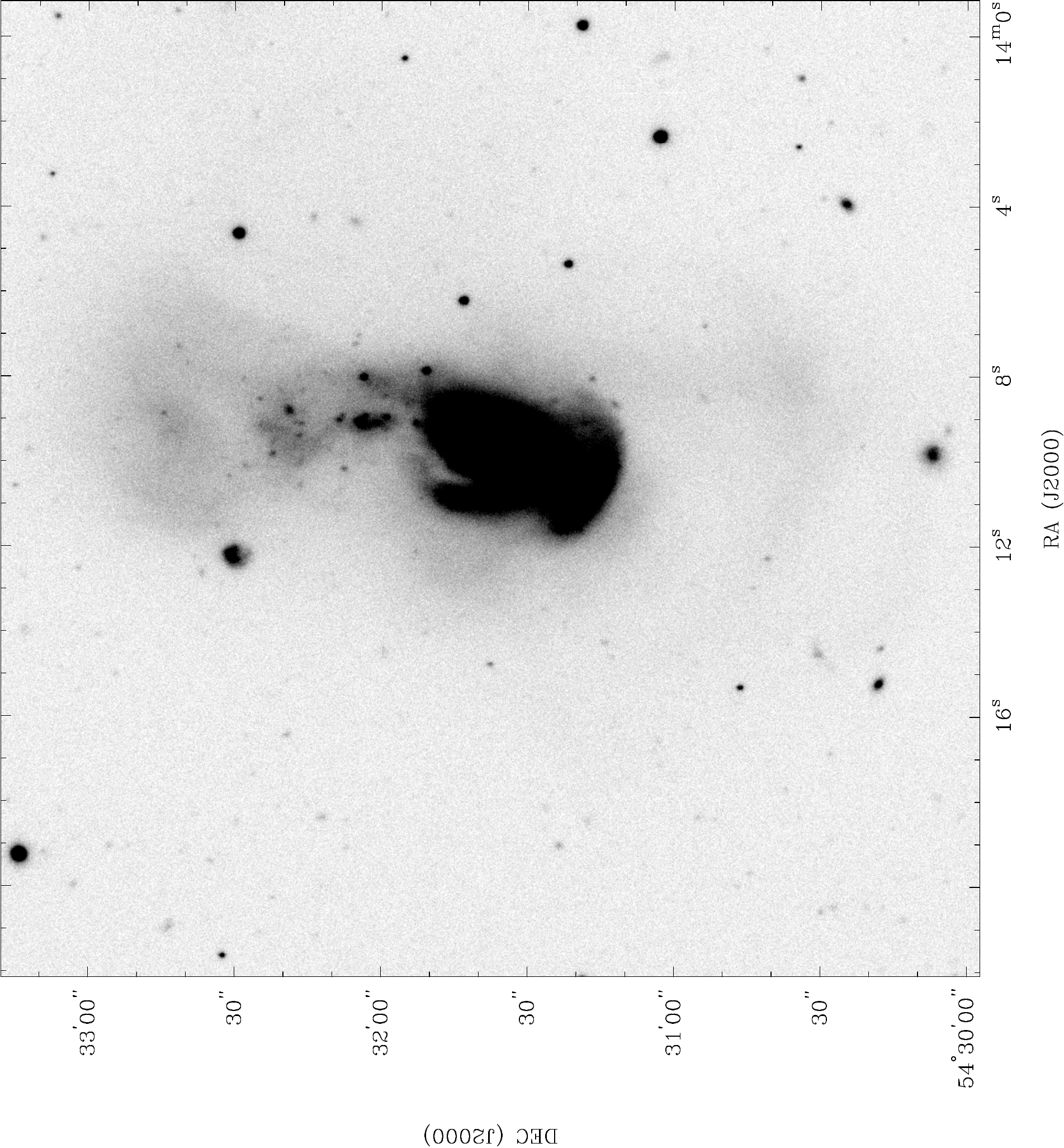}
   \includegraphics[angle=270,width=8.5cm]{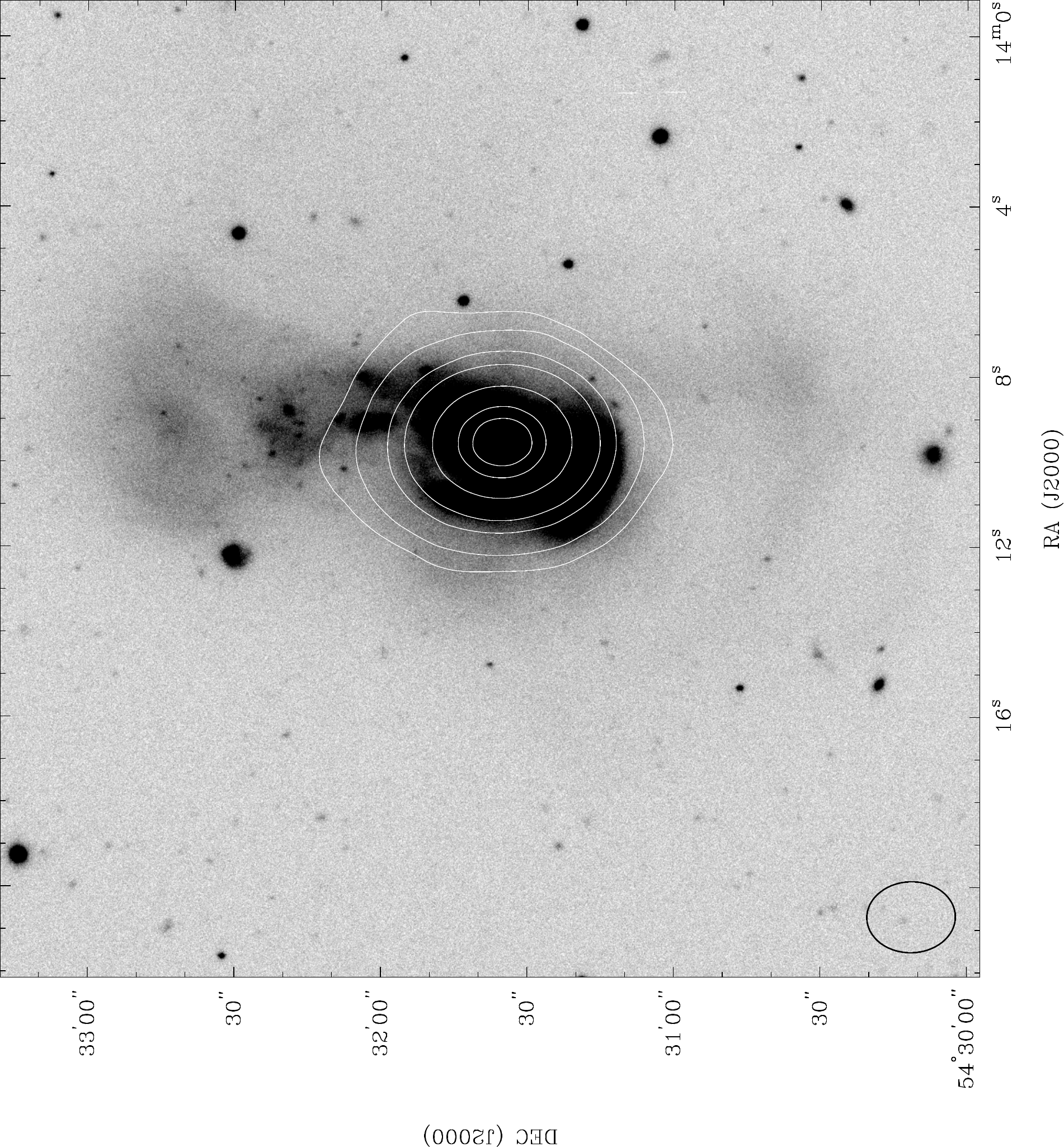}
   \caption[Optical image and 20\,cm continuum contour map of NGC\,4194]
{Left: Nordic Optical Telescope V-band image of the Medusa. 
Right: Contour map of
              the 20\,cm continuum emission superimposed on the V-band image. The 
              synthesised beam is shown in the lower left corner.
              The contour 
              levels are 0.2, 0.5, 2, 5, 20, 50, 70\,mJy\ beam$^{-1}$. 
              }
         \label{n4194chap_contimage}
   \end{figure*}

The Medusa galaxy (NCG\,4194, Arp~160, IZw33) is a nearby galaxy ($D=39$~Mpc) with the
 disturbed appearance typical of a merger (see Fig.~\ref{n4194chap_contimage}). 
It has been interpreted as a 
 candidate for a minor merger between a large 
elliptical and a small spiral galaxy, a so-called S+E merger 
\citep{2000A&A...362...42A}. 
The galaxy earned its nickname from the fuzzy and knotty optical tail to the
 North, 
 reminiscent of hair 
 or tentacles. A sharp shell-like structure is
 visible on the opposite side of the galaxy, confining the optical main body. A second,
 fainter shell 
can be seen further out 
on deep images (e.g.,
 \citealt{optsample}).

Major mergers
between two gas-rich disk galaxies 
can 
lead to ultraluminous infrared
galaxies (ULIRGs, $L_{FIR} > 10^{12}\,L_{\odot}$) due to an extreme increase
of star formation \citep{2003AJ....126.1607S}.
Strong gas inflows are triggered in these mergers, increasing the gas density in 
the centre and thus leading to the observed starburst 
\citep[e.g.,][]{1991ApJ...370L..65B,1996ApJ...471..115B,1996ApJ...464..641M}. 
In these simulations the merger remnant evolves into a regular elliptical galaxy.
In contrast,
 in minor, i.e., unequal-mass
mergers 
only a moderate, if at all, starburst is induced implying a 
lower far-infrared (FIR) luminosity (less than 
 $10^{11}\,L_{\odot}$) \citep{2008MNRAS.384..386C}. 
 The triggered star formation and the remnant galaxy type strongly depend on the
mass ratio of the merging pair, but not so much on the total gas mass 
\citep{2005A&A...437...69B,2008MNRAS.384..386C}.  In contrast to most nearby minor
mergers, in which the major galaxy is a spiral and thus dominating the star formation
(e.g., M\,51),
the Medusa gives us an ideal opportunity to investigate the influence of the
merger on the smaller companion because it is the only gas-rich partner here. 
Since the merger-induced star formation does
not depend on the gas content of the companion \citep{2008MNRAS.385L..38M}, the results
of the Medusa might be applicable to other minor mergers, independent on the type
of the large partner. 
\\
Another significant 
difference between ULIRGs and 
galaxies like the Medusa is 
the  
extent of the starburst induced by the merger. 
While the burst region in ULIRGs is
compact and typically concentrated to the inner kiloparsec, 
(e.g., \citealt{1996ARA&A..34..749S}), the
Medusa exhibits an extended region of ongoing star formation which is,
 however,
 not as
intense as in ULIRGs
(\citealt{1990ApJ...364..471A}, \citealt{1994ApJ...422...73P}).
 \cite{2004AJ....127.1360W} analyzed optical and ultraviolet Hubble Space Telescope (HST) images of the
 nucleus and found 
 star-forming knots younger than 20\,Myr in the center. They derived a star
 formation 
 rate (SFR) of $\rm \sim 6\,M_{\odot}\ yr^{-1}$, but argued that the
 overall  SFR
in the Medusa might be as high as $\rm 30\,M_{\odot}\ yr^{-1}$ since the knots produce
 only 20\% of the UV flux in the center.
Based on recent HST spectroscopic observations,
 \cite{2006AJ....131..282H} estimate an even higher total SFR of $\rm \sim
46\,M_{\odot}\ yr^{-1}$, and ages of individual star forming regions of
5.5--10.5\,Myr. 

The Medusa possesses a reservoir of gas, both molecular and atomic. 
High resolution aperture synthesis maps 
obtained with the Owens Valley Radio Observatory 
revealed extended CO emission 
on a total scale of 25$''$ (4.7~kpc) \citep{2000A&A...362...42A}. 
The CO morphology is complex, occupying 
mainly the center and the north-eastern part of the main optical body.
The extended CO gas follows two prominent dust lanes:
one which is crossing the central region at right angle with
respect to the optical major axis, 
and a second which curves to the north-east and
into the beginning of the northern optical tail. 
\cite{2000A&A...362...42A} 
suggested 
that the central starburst is being fueled by gas flows along the central dust
lane.
The presence of multiple spectral peaks ($\Delta$V $\approx$ 200 \kms) 
suggests that at least two 
velocity systems coexist in the CO emitting gas. 
Either this is an effect of substantial warping of the main body, 
or simply the unrelaxed, overlapping
systems of the progenitor galaxies. 
The multiple features seem to be related to the
extended dust lanes, but 
several velocity components co-exist also at the base of the optical tail. 
Single-dish measurements revealed the presence CO emission even
 inside  the optical tail out to 4.7\,kpc away from the center  
 \citep{2001A&A...372L..29A} where  there is no sign of ongoing star
 formation \citep{1990ApJ...364..471A}. The total molecular
 gas mass in the inner 2 kpc was estimated to $2\cdot 10^9\,M_{\odot}$. 
Thus,
the star formation efficiency is almost as high as in ULIRGs although the
molecular gas density is significantly lower. 
\citep{2000A&A...362...42A}. Besides uncertainties in the molecular gas 
conversion factor, this gives a hint that the ISM in these mergers is
different compared to the molecular cloud complexes in normal galaxies and also
in extreme starbursts. \cite{2000A&A...362...42A} argue that 
the observed solid-body rotation may offer a shear-free environment. This
may support cloud collapse and thus star formation. The star formation rate
might have increased because a large amount of molecular gas was transported
to the centre on short time scales by the interaction. 
%----------------------------------------------------------------------------
%----------------------------------------------------------------------------

The galaxy contains a similar amount of atomic gas as molecular gas (\mhi $\sim 2.2 \cdot 10^9\,M_{\odot}$), 
inferred from the single-dish \hi emission spectrum obtained by \cite{1981ApJ...247..823T} using
the Greenbank 91\,m telescope.
Atomic gas has recently been detected in absorption toward the center of the Medusa, 
through sub-arcsecond MERLIN observations 
\citep{2005A&A...444..791B}. 
The kinematics of the \hi gas 
in the central region of the galaxy 
is similar to that of the molecular gas traced by the 
CO emission
\citep{2000A&A...362...42A}, 
which suggests that the atomic and molecular
 component of the cold 
interstellar medium 
are probably physically related.
Two compact radio sources were found in the central parsec,
embedded in a region of diffuse radio continuum emission. Even though the 
presence of a weak AGN cannot be ruled out completely, most (if not all) of the continuum emission
is related to the starburst \citep{2005A&A...444..791B}. 

NGC~4194 displays conflicting evidence about its merger age. 
The sub-arcsecond resolution MERLIN data revealed two compact radio
components in the center, separated by $\rm \sim 0^{''}.35$,
corresponding to 65\,pc only \citep{2005A&A...444..791B}. The optical
tail is fairly diffuse, 
which is suggestive of an 
advanced merger where the progenitors have coalesced. 
In contrast, the fairly widespread 
CO distribution, irregular optical isophotes, multiple spectral CO features, 
and the distorted central
body hint toward a younger system.

In this paper, we present observations and $N$-body numerical simulations 
of the stars and the gas in the Medusa merger. 
We mapped the large-scale distribution and kinematics of the \hi gas  
with the Westerbork Radio Synthesis Telescope (WSRT) 
and used the Nordic Optical Telescope (NOT) to obtain 
a deep V-band image of the distribution of the stars in the Medusa. 
Then we carried out numerical simulations which we compared to the observations. 
\hi 21 cm line observations are a powerful tool 
to trace the dynamical history of interacting and merging galaxies 
and provide constraints to dynamical models of 
the history of the interaction. 
They complement CO millimeter-wave line observations
that trace mainly the inner part of galaxies where the molecular gas resides. 
Furthermore, they offer is the only way
to trace the distribution and kinematics of gas in large scale tidal features. 
The observations are presented in section~2 and 3, and the simulations in section~4. 
Follows a discussion. 

\begin{table}
      \caption[Basic properties of NGC\,4194]{Basic properties of
	NGC\,4194. The distance is based on ${\rm 
	H_0 = 75\,km\ s^{-1}\ Mpc^{-1}}$.
	}
        \label{n4194chap_intro}
\centering
\begin{tabular}{lc}
\hline\hline
RA (2000) & 12 14 09.5\\
DEC (2000) & +54 31 37 \\
$\rm v_{hel,opt}$ (km\ s$^{-1}$) 		& 2442 \\
D (Mpc) 				& 39\\
$\rm L_B$ ($10^9\,L_{\odot}$) 		& 15.7\\
$\rm L_{FIR}$ ($10^{10}\,L_{\odot}$) 	& 8.5\\ 
FIR flux at 60\,$\rm \mu$m (Jy) 	& 25.66\\
FIR flux at 100\,$\rm \mu$m (Jy) 	& 26.21\\
\hline

 \hline
\end{tabular}
\end{table}

%---------------------------------------------------------------------------------------
\section{Observations and data reduction}
\subsection{\hi observations}

\begin{table}
\caption[WSRT \hi observations of NGC\,4194]{Observing parameters of the WSRT
  observations.} 
\label{n4194chap_wsrtobs}
\centering
\begin{tabular}{lc}
\hline\hline
 observing parameters  &     \\
\hline
date 			&12.7.2003  \\
central frequency 	& 1408.6 MHz\\
total bandwidth		& 20\,MHz \\ 
number of channels	& 256 \\  
velocity resolution 	& 4.12 \kms \\   %97.656kHz
synthesised beamwidth	& $23'' \times 19''$\\
primary beam 		& $ 36\,'$\\
primary calibrator	& 3C147 \\
\hline
\end{tabular}
\end{table}
%---------------------------------------------------------------------------------------
%---------------------------------------------------------------------------
%N4194 large % FIG2 
 \begin{figure*}
   \centering
   \includegraphics[angle=270,width=\textwidth]{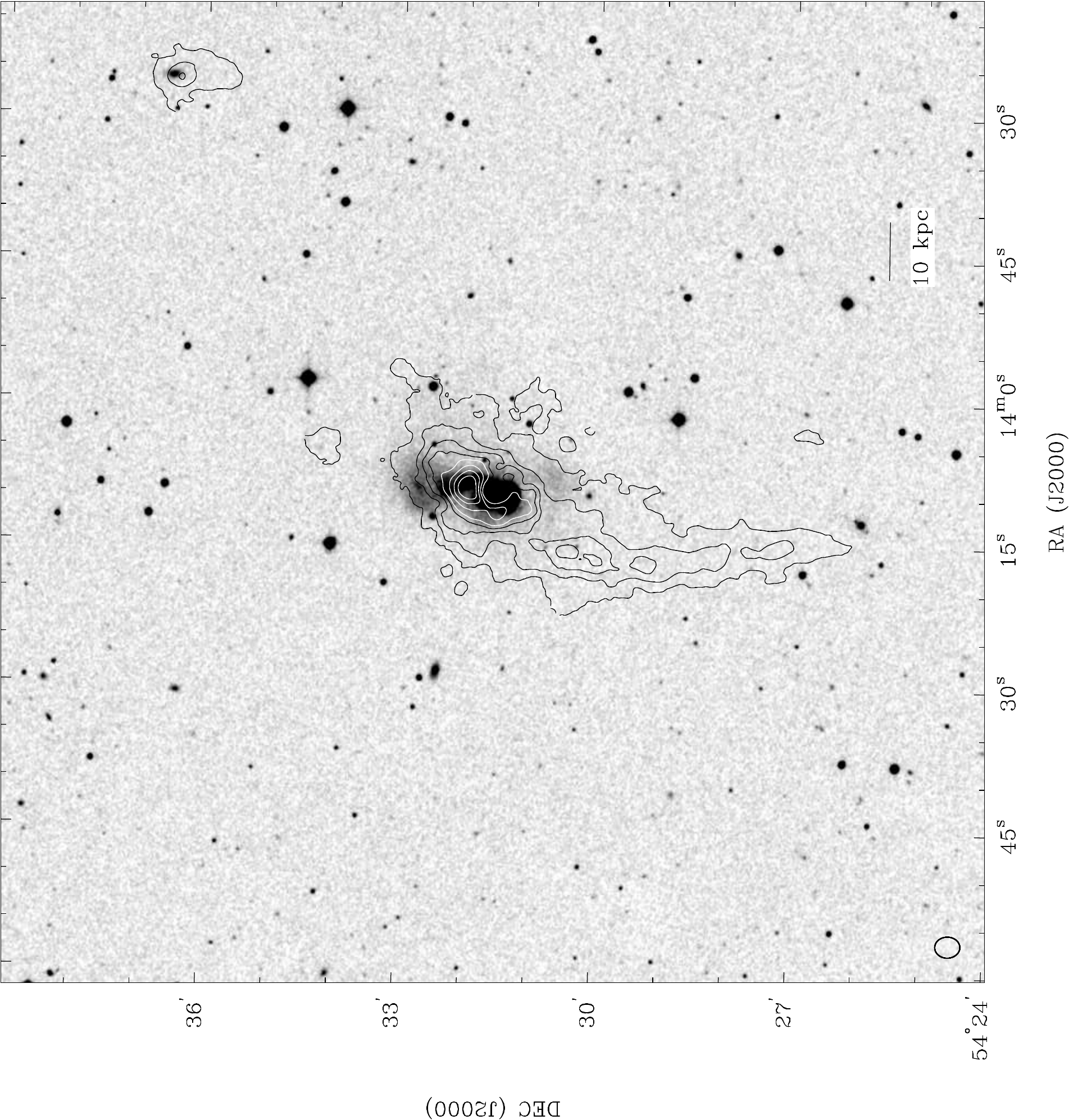}
      \caption[NGC\,4194: Surroundings in \hi]{The \hi distribution as contours
	overlayed on an optical image of the field around
	NGC\,4194 from the Digitized Sky Survey. The contour levels are
 	0.01,0.05,0.1,0.15,0.2,0.3,0.4,0.5\,Jy\ beam km\ s$^{-1}$. 
	The bar in the lower right corner marks a projected distance of
	10\,kpc, assuming a distance of 39\,Mpc. The lack of contours at the
	center of NGC\,4194 is due to an absorption source. The beam is shown in the
	lower left corner. 
            }
         \label{n4194chap_n4194large}
   \end{figure*}
%---------------------------------------------------------------------------

%------------------------------------------------------------------------------------
%NGC4194: HI spectra (also of companion) FIG 3
 \begin{figure*}
   \centering
   \includegraphics[angle=0,width=8.8cm]{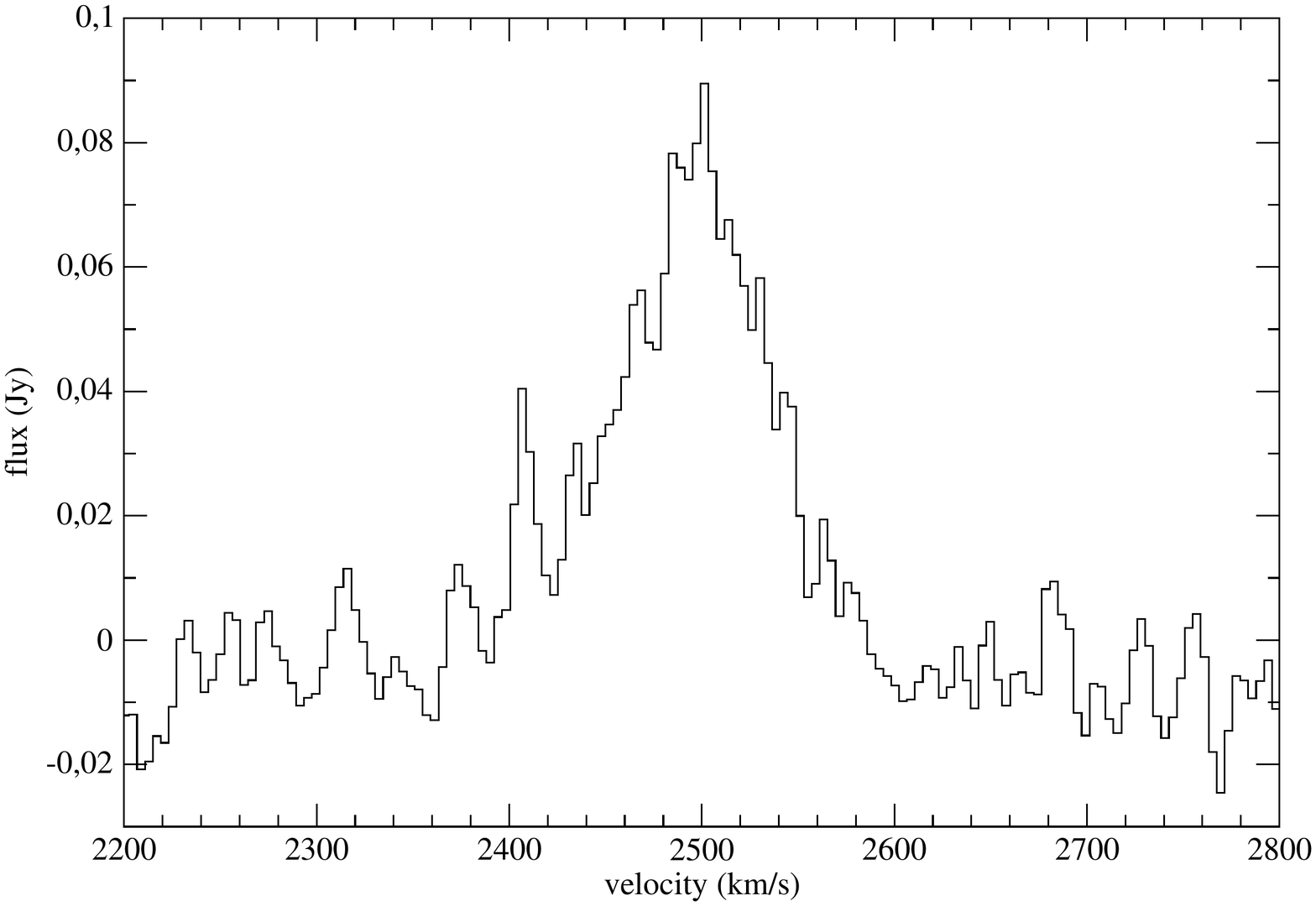}
    \includegraphics[angle=0,width=8.8cm]{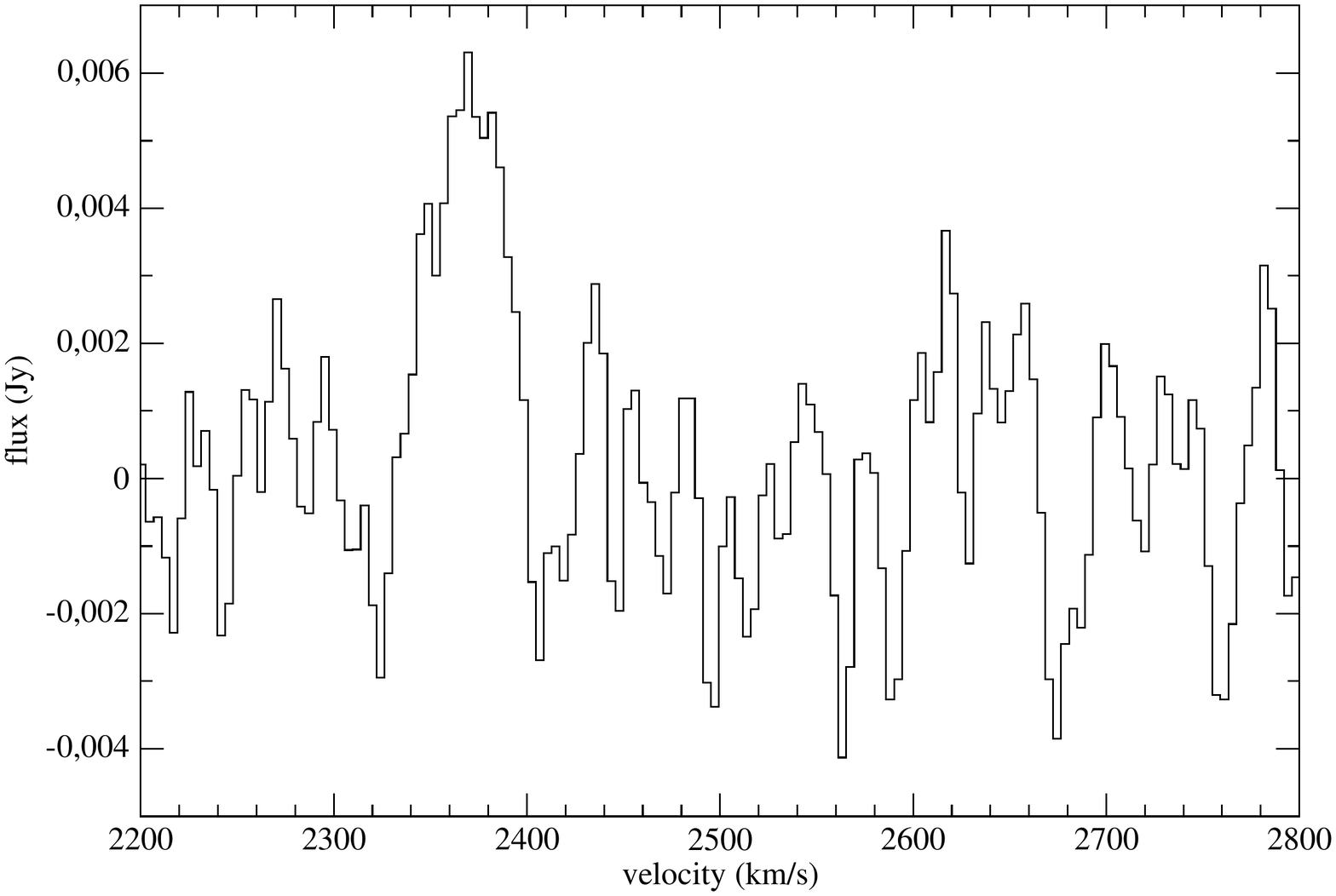}
      \caption[NGC\,4194: \hi spectra]{Integrated \hi spectra of NGC\,4194 (left)
              and the companion dwarf galaxy to the North-West 
              (right).}
         \label{n4194chap_n4194hispec}
  \end{figure*}
%------------------------------------------------------------------------------------

%-------------------------------------------------------------------------------------------------
%NGC4194: mom0 & cont, vel. & cont., optisch & mom0 overlay, mom2 & cont.
% FIG 4
 \begin{figure*}
   \centering
   \includegraphics[angle=270,width=6.6cm]{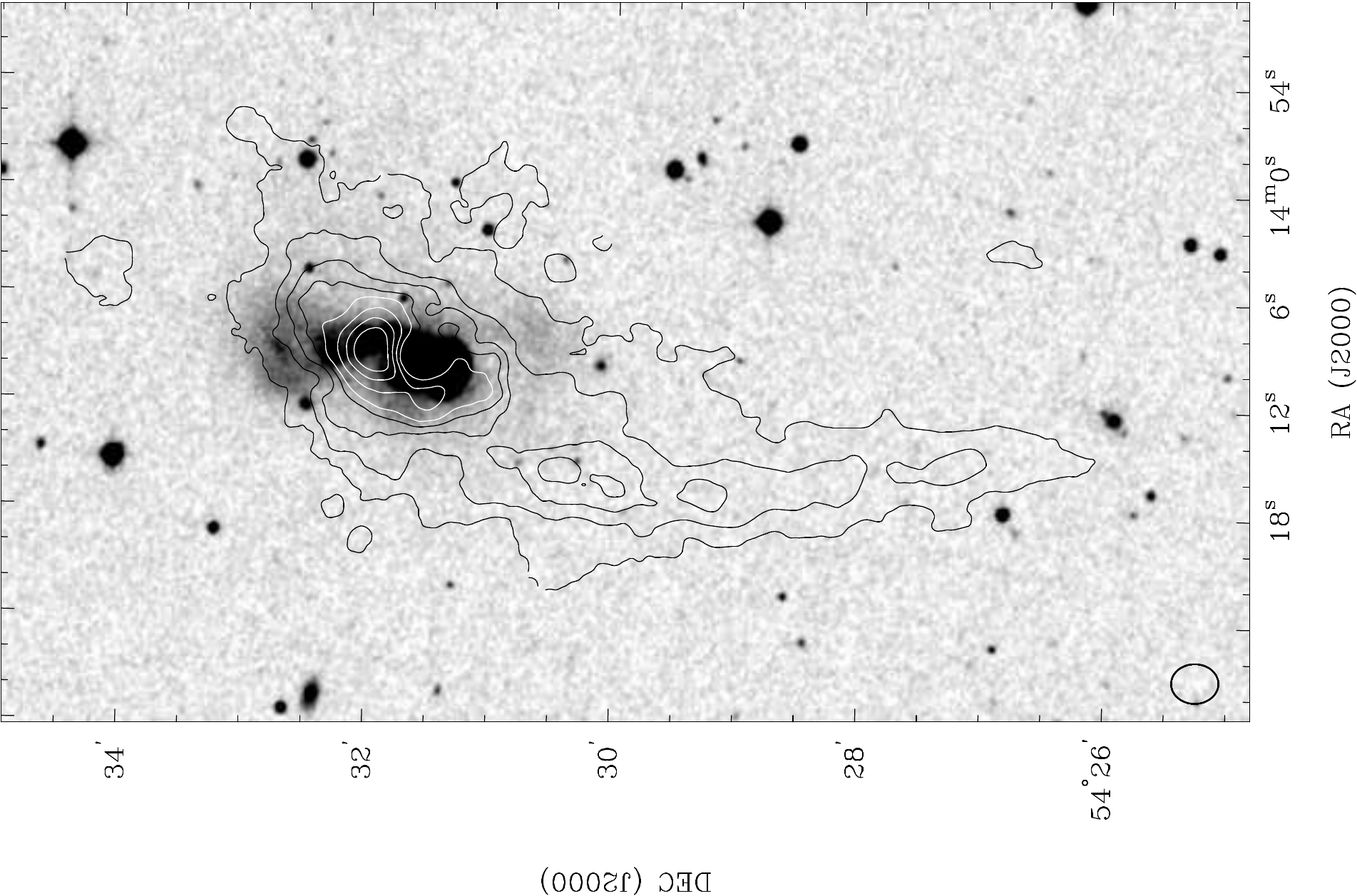}
   \includegraphics[angle=270,width=7.8cm]{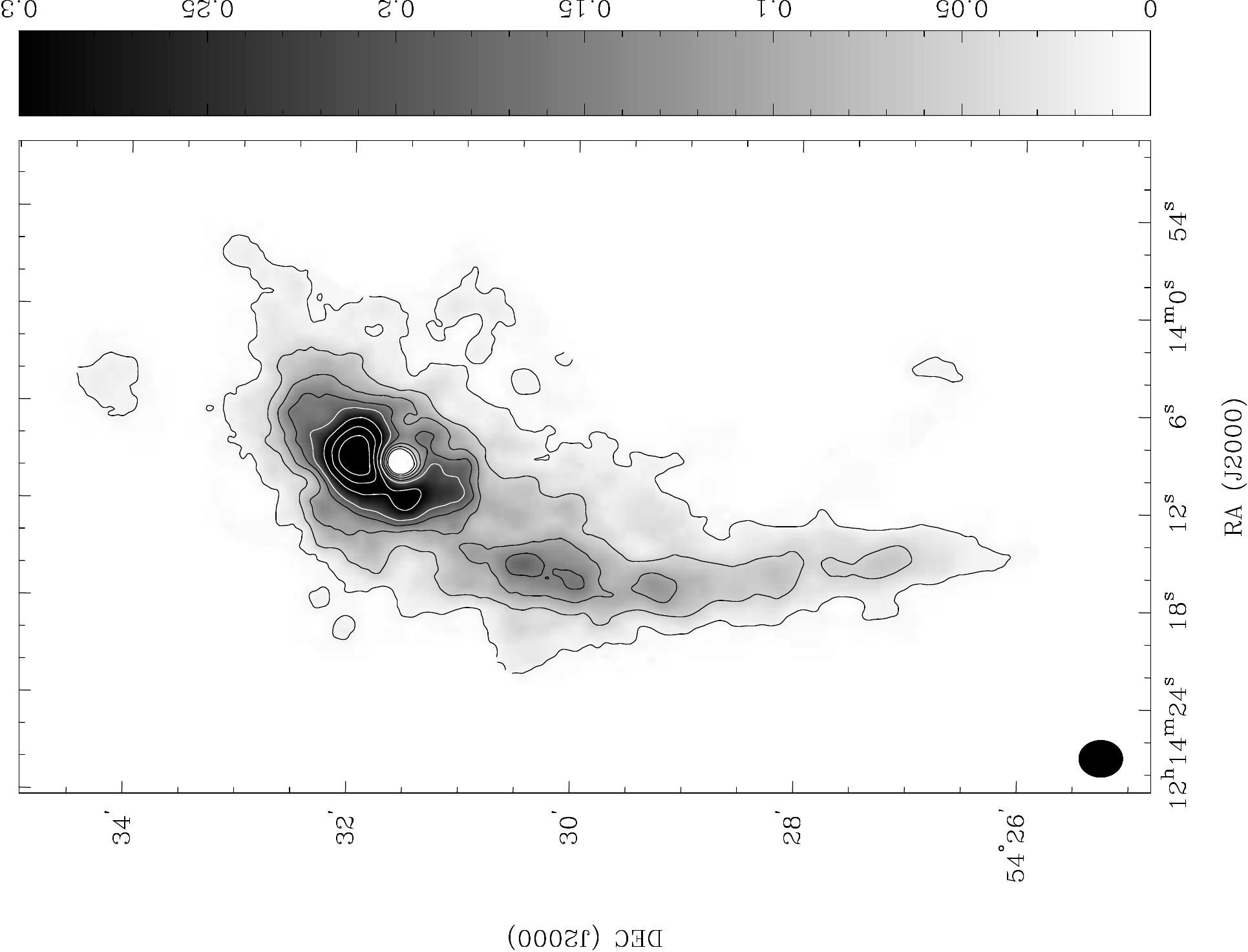}\\
   \includegraphics[angle=270,width=7.8cm]{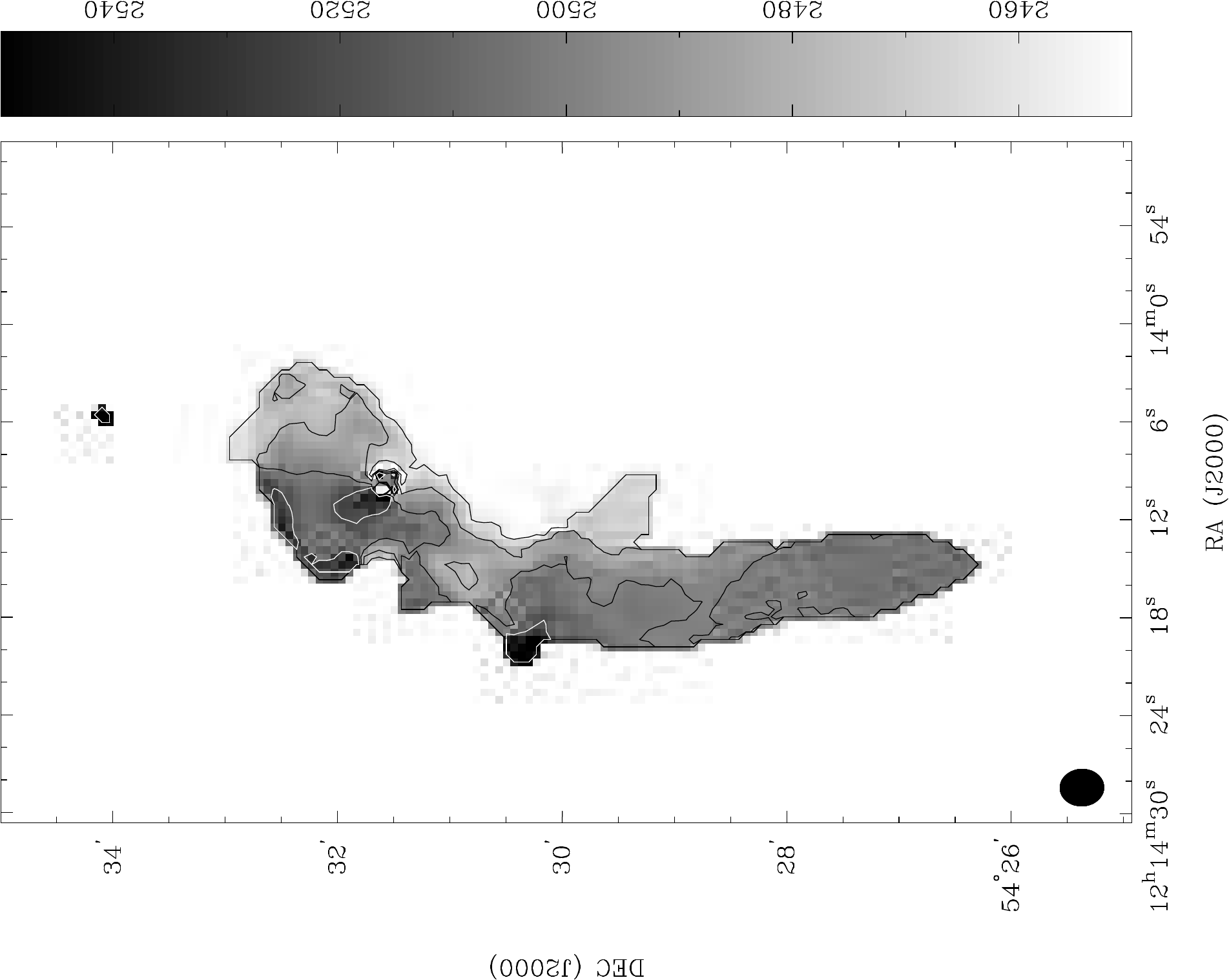}
   \includegraphics[angle=270,width=7.5cm]{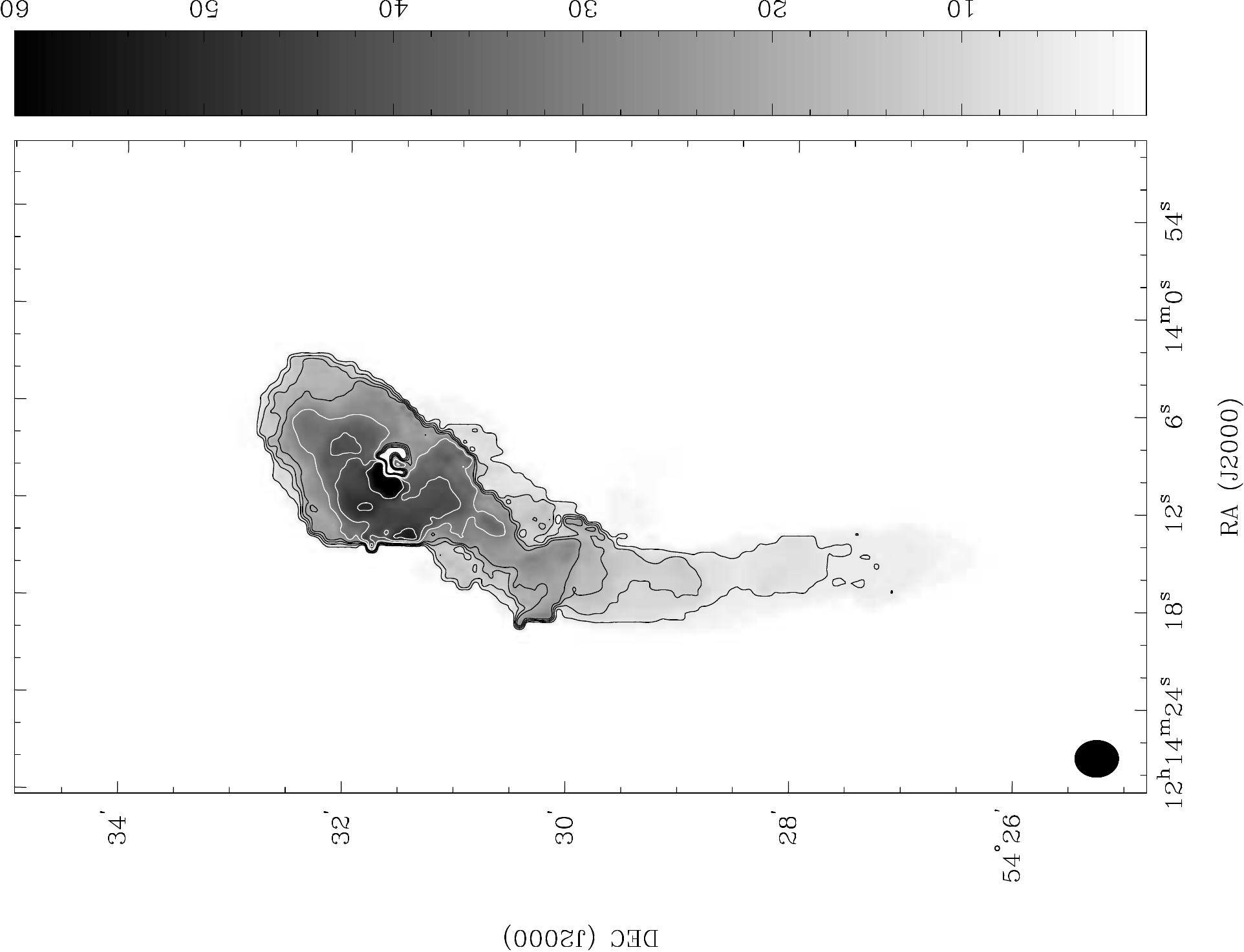}
      \caption[NGC\,4194: \hi data]{Maps of the \hi distribution, velocity field and 
velocity dispersion. (top left) 
              \hi distribution
	      overlayed 
	      on optical image from the Digitized Sky Survey. The contour levels are
              0.01,0.05,0.01,0.15,0.2,0.3,0.4,0.5\,Jy\ beam\ km\ s$^{-1}$
              (0.01\,Jy/beam\ km/s corresponds to $\rm 2.47\cdot
              10^{19}\,cm^{-2}$),  . 
              (top right)
              \hi distribution as map and contours. Levels are the same. The
              white dot marks the region where \hi absorption is found.
              (bottom left)  
              Velocity field, contour levels are from 2460\,km\ s$^{-1}$ to 2600\,km\ s$^{-1}$
	      in steps of 20\,km\ s$^{-1}$. 
	      (bottom right) 
              \hi velocity dispersion, contour
              levels are 5, 10, 15, 20, 30, 40, 50\,km\ s$^{-1}$.
	      The synthesised beam is marked in the bottom left
              corner.  
              }
         \label{n4194chap_n4194hi}
  \end{figure*}
%--------------------------------------------------------------------------------
%NGC4194: channelmap FIG 5 
 \begin{figure*}
   \centering
   \includegraphics[angle=-90,width=15cm]{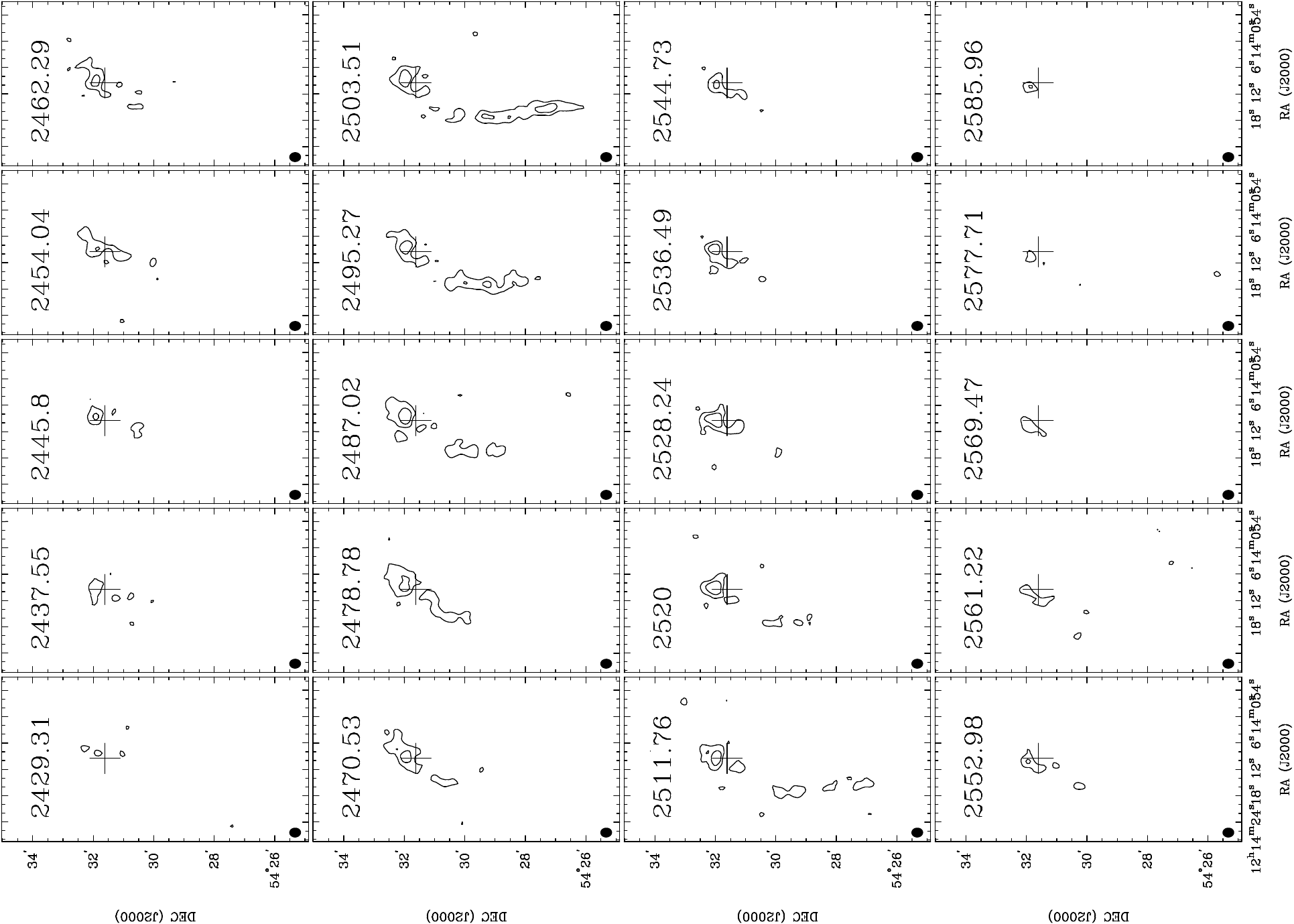}
      \caption[NGC\,4194: \hi channel maps]{\hi channel maps. 
              The optical 
              center position is marked with a cross. The contour levels are
              $\rm 1, 2, 5 \times 0.0016\,Jy\ beam^{-1}$. 
		The velocity in the
		Local Standard of Rest 
              is given at the top, the beam is shown at the bottom left corner. 
              }
         \label{n4194chap_n4194chanmap}
  \end{figure*}
%--------------------------------------------------------------------------------

The observations were
carried out with the WSRT in July 2003. We observed one full 12h track in the maxi--short
array configuration under good
conditions. As a flux calibrator, the standard source 3C147 was observed
before and after the target. No phase calibrator was observed,
since the standard procedure at WSRT for phase calibration is to obtain
self-calibration. In Tab.\,\ref{n4194chap_wsrtobs} we list the observing
parameters and 
give information about the setup used. We chose parameters
which allowed us to both cover a large velocity range, because \hi lines can
typically be broader in mergers than in normal disk galaxies, and to reach a
high resolution in velocity to be able to detect narrow features.

The data were reduced using {\tt MIRIAD} together with additional, WSRT
 specific routines written by T. Oosterloo\footnote{http://www.astron.nl/$\sim$oosterlo/wsrtMiriad/}. The quality of the data was good,
 so that only little flagging of interferences was necessary. 
First, the data had to be calibrated to system temperature scale. Since no
secondary calibrator was observed, after flux and bandpass calibration using
the primary calibrator, a 
continuum image was created to perform self-calibration on continuum sources
within the field. 
To achieve a deep CLEANing, we iteratively created a mask which includes
 emission only and which defines the regions to be CLEANed. After CLEANing and
 self-calibration a new mask was created and the process repeated until
no improvement could be made.  The results were then applied to the
target galaxies.

CLEANing the \hi cube was done in a similar way to CLEANing the
continuum. After 
continuum
subtraction, we applied Hanning smoothing, since this was not done during the
observations.
Again, we used as mask containing emission only to define CLEANing boxes.
This
 procedure was repeated 3 times until the noise level was reached.
 Integrated intensity and velocity maps were created using the {\tt MOMENT}
task in {\tt MIRIAD}. Finally, primary beam correction was applied to all data.
% A more detailed description of the reduction procedures will
%be 
%given in \cite{hipaper}. 

We produced two sets of maps: naturally weighted ones
(with an rms noise level of 0.4\,mJy\ beam$^{-1}$, resolution of
$31''\times27''$), and maps obtained using a  
{\tt ROBUST} factor of 0.5, which gives a better resolution with slightly less
sensitivity (noise level of 0.6\,mJy\ beam$^{-1}$, resolution of 
23\arcsec$\times$ 19\arcsec). 
Since the morphological features were the same in 
both sets of maps, 
we present here only the {\tt ROBUST 0.5} maps 
in which  
structures in the \hi tail 
appear more clearly. 

\subsection{Optical imaging}
We also present here optical Johnson V-band images taken with ALFOSC mounted
at the 
Nordic Optical Telescope (NOT) on La Palma, Canary Islands in January
2003. The field of view 
of ALFOSC is $6\farcm4\times 6\farcm4$, thus large enough to cover not only
NGC\,4194 completely but also a sufficient region of sky for a proper
background 
subtraction. The integration time was 10 minutes. Standard reduction was
applied to the data using the IRAF software. A further description of
the observations and reduction as well as a detailed analysis of the optical
data will be given in \cite{optsample}.

\section{Observational results}

\subsection{Radio continuum}
Figure~\ref{n4194chap_contimage} shows the NOT V-band image of NGC\,4194 
on which contours of the 20\,cm radio continuum emission have been 
superimposed.  
The flux density of the source is 0.108\,Jy. We
see continuum emission extending out to a deconvolved size of $6\farcm6$
(1.2\,kpc)  
associated with the galactic main
body. 
 This is
not surprising, since the 20\,cm continuum flux is a tracer of star
formation and NGC\,4194 exhibits an extended region of
intense star formation. The continuum map shows a smooth distribution center
at the position of the optical nucleus. 

\subsection{Neutral hydrogen}
Figure~\ref{n4194chap_n4194large} shows the
integrated \hi intensity map overlayed on an optical image 
from the Digitized Sky Survey (DSS). 
\hi gas is found all over the optical extent of the Medusa. 
Remarkable is
the single long \hi tail to the South. 
\hi is also found in absorption toward the center. 
We also detect \hi associated with the small nearby galaxy SDSS~J121326.03+543631.7
identified in the 
Sloan Digitized Sky Survey about $8\arcmin$ to the north-west. 

The \hi properties of NGC\,4194 and of the nearby dwarf galaxy are summarised in
Tab.~\ref{n4194chap_himass}. 
% ch
The integrated \hi spectra 
are shown in Fig.~\ref{n4194chap_n4194hispec}.
We calculate a total \hi mass of NGC\,4194 of $\rm 2
\cdot 10^9\,M_{\odot}$, in good agreement with the \hi mass
determination by \cite{1981ApJ...247..823T} of $\rm 2.2  
\cdot 10^9\,M_{\odot}$. This shows that 
our interferometric observations do not miss any extended 
emission.

\begin{table}
\caption[\hi properties of NGC\,4194]{
\hi properties of the Medusa merger and of the nearby 
dwarf galaxy  SDSS~J121326.03+543631.7. 
The values listed for
  NGC\,4194 include the contribution from the \hi tail.}
\label{n4194chap_himass}
\centering
\begin{tabular}{lccc}
\hline\hline
 \hi properties & NGC\,4194 & Tail & Nearby dwarf\\
\hline
 ${\rm v_{cent}}$ (km/s)& 2500& & 2370\\
 range (km\ s$^{-1}$) & 2400 -- 2580& 2470 -- 2520& 2350 -- 2390\\  % 190 km/s
 extension (kpc) & 81 & 56 & 14 \\
 ${\rm F_{HI}}$ (Jy\,km\ s$^{-1}$) & 5.65& 2.03& 0.20\\
\mhi ($10^9$\,\msun) & 2.0& 0.7&0.07 \\
\mhi / \lb & 0.18& --& --\\
\hline
\end{tabular}
\end{table}

\subsubsection{The main body of NGC~4194}

\begin{figure}
   \centering
   \includegraphics[angle=270,width=8.8cm]{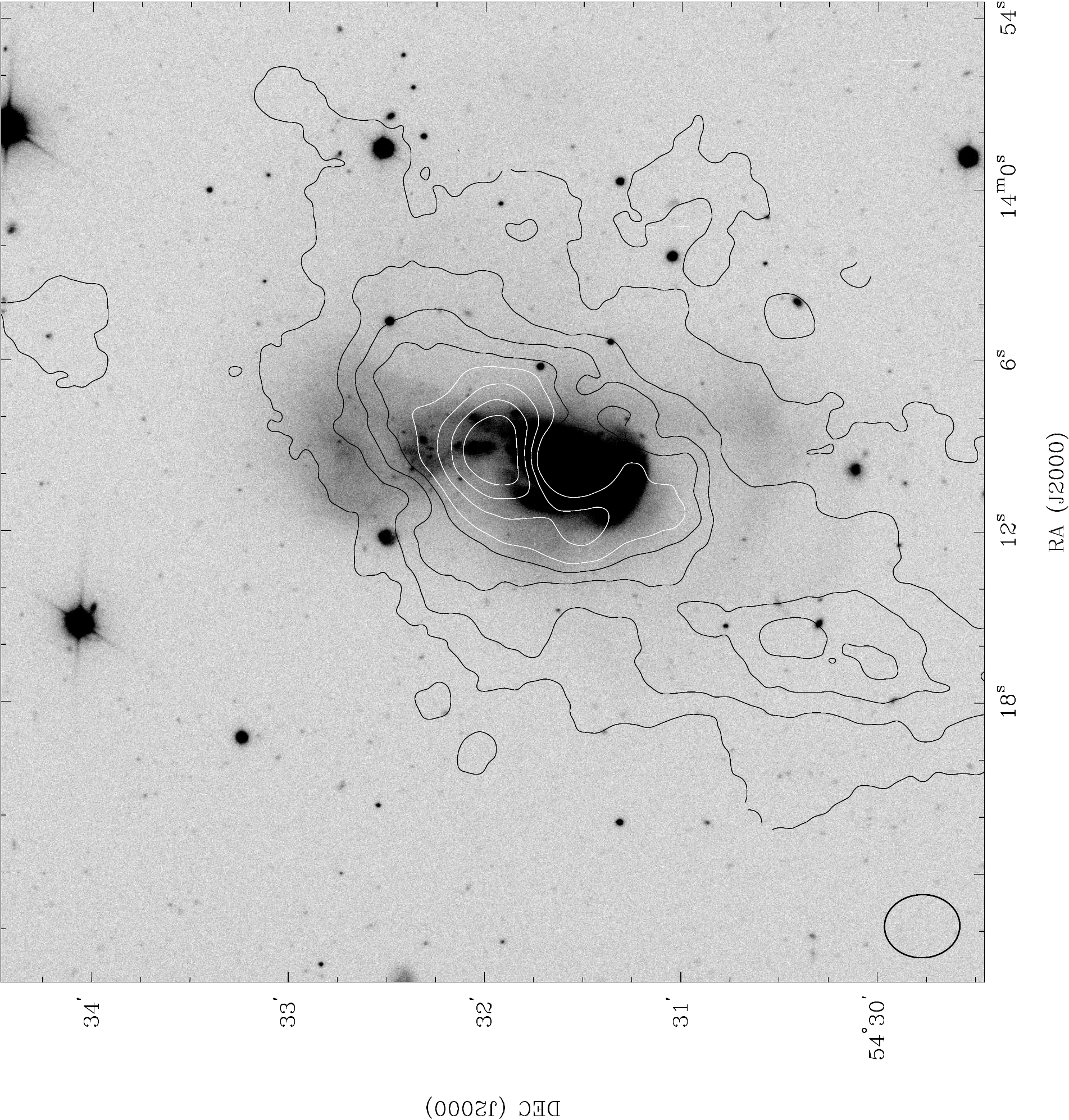}
      \caption[NGC\,4194: \hi map of the main body] {Contours of the \hi intensity map overlayed on
            the V-band image. 
The optical
            image covers only the upper part of the \hi tail. The contour levels are
            0.01, 0.05, 0.1, 0.15, 0.2, 0.3, 0.4, 0.5\,Jy\ beam$^{-1}$ km\ 
s$^{-1}$.
            }
         \label{n4194chap_notimage}
   \end{figure}

Figure~\ref{n4194chap_n4194hi} shows the \hi distribution, 
the velocity field 
and the velocity dispersion map. 
The channel maps are shown in Fig.\,\ref{n4194chap_n4194chanmap}.
Figure~\ref{n4194chap_notimage} shows the \hi integrated emission as isocontours 
overlaid on our optical V-band image, which is deeper than the DSS image 
but covers only the upper part of the \hi tail. 
The \hi emission peaks north of
the absorption feature, aligned with the beginning of the optical tidal tail
and some bright knots, which might be proto-star clusters.
The \hi  embeds
the optical tail completely, but is not considerably extended further out,
i.e., no pronounced \hi tail as a counterpart of the optical tail is found.
The optical tail seems to be embedded in a smooth \hi distribution
somewhat extended towards the north-west, maybe coincidentally in the direction
where the possible companion is found.

In the optical, we see two shells in the south, a bright one close to the
center 
and a 
second, fainter one at a larger radius. We do not see any feature, neither in
the \hi  
distribution nor in the velocity and velocity dispersion maps which is related
to these shells. 

The velocity map (Fig.~4, bottom left) shows a slightly disturbed rotating disk,
represented by a typical spider diagram, with regions of higher velocity on
the eastern side and lower velocity on the western side on the main body
of the Medusa. 

The \hi velocity dispersion is larger in the galactic body than in the
tail. The highest values are measured north-east of the absorption region,
where there is a slight dip in the \hi distribution. This region is also
roughly related to the dust lane where extended molecular gas was found 
\citep{2000A&A...362...42A}.  In the molecular gas maps multiple velocity
components are
found, mimicing high velocity dispersion in an integrated dispersion map. It seems
therefore likely that also the high dispersion in \hi is related to the dust lane
rather than the starburst region, even 
though the spatial resolution of the \hi presented here
is not sufficient for a detailed 
investigation.

\subsubsection{\hi in absorption toward the center}
\begin{figure}
   \centering
   \includegraphics[angle=0,width=8.8cm]{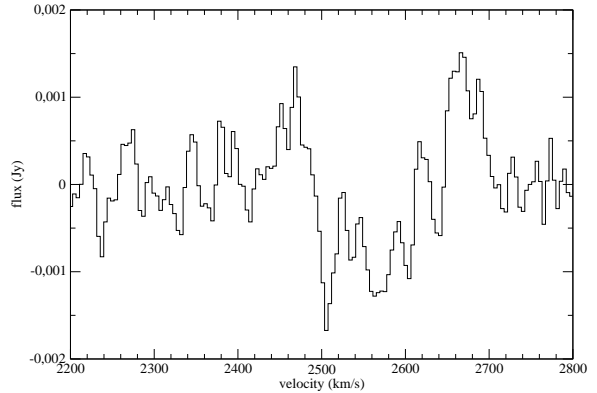}
      \caption[NGC\,4194: \hi absorption spectrum]{ The \hi absorption spectrum
toward the center 
            of NGC\,4194. 
            }
         \label{n4194chap_absorption}
   \end{figure}

\hi is seen 
in absorption against the underlying continuum source of the star forming
nucleus. Fig.~\ref{n4194chap_absorption} 
shows the measured spectrum. 
The multicomponent velocity structure clearly seen in the MERLIN data and
discussed by \cite{2005A&A...444..791B} is tentatively seen here, too.
Our spectrum
shows at least two components, one at $\rm \sim 2500\,km\ s^{-1}$ and one at 
\linebreak
$\rm
\sim 2570\,km\ s^{-1}$.

\subsubsection{The \hi tail}
We detected a
single
remarkable \hi  tail going to the South out to $\sim$56\,kpc from the
center. 
The \hi tail contains several clumps embedded in smooth, diffuse \hi 
emission. 
Neither in our V-band image 
(which, unfortunately, covers only the upper part of the \hi tail)
nor
in the  
larger DSS image  
did we find any optical counterpart 
of the \hi clumps.
We determined the \hi mass in the tail as $\rm \sim 7\cdot10^8\,M_{\odot}$,
i.e. 35\% of the 
total \hi mass. 
The velocity in the
tail is smoothly increasing, with the highest velocity of $\rm \sim
2510\,km\ s^{-1}$ at the tip of the tail. 
In the velocity dispersion map the smooth behaviour of the tail is found as a
continuous but not disturbed increase of dispersion from the tip of the tail
towards the galactic body. The dispersion is generally higher at positions of
the clumps, especially in two large clumps close to the galaxy's main body.\\
The lowest velocity ($\rm \sim 2455 - 2480\,km\ s^{-1}$) is
found in the region between the central \hi concentration and these two bright
clumps, marking the beginning of the tail. In this region, there is a 
gradient in velocity from north-east to south-west, ending in an area having
the 
lowest velocity in the system southwards of the center. North-west of the two
clumps the smooth decrease of velocity is intercepted by a 'ridge' of
slightly increased velocity. In the velocity dispersion map we see enhanced
dispersion there ($\rm \sim 35\,km\ s^{-1}$ compared to $\rm \sim 10\,km\ s^{-1}$ left
and right of this 'ridge'), whereas in the \hi distribution map, if at all,
there seems to be a slight dip in \hi intensity associated with this 'ridge'.

\subsubsection{\hi in a nearby dwarf galaxy}
We were also able to detect \hi in the dwarf galaxy J121326.03+543631.7 at a
distance 
of $8.06'$ (91\,kpc) to the North-West ($\rm \alpha= 12^h 13^m 26.38^s,
\delta= +54^{\circ} 36' 31''$). Until now, it was not known that
NGC\,4194 has a companion: our \hi measurements show that this dwarf has a
central velocity of 2380\,km\ s$^{-1}$, which is only $\sim$ 120\,km\ s$^{-1}$ lower than that
of NGC\,4194, 
thus 
probably indicating that the two objects are physically
connected. The measured \hi velocity is in good agreement with its
optical velocity of 2436\,km\ s$^{-1}$ (SDSS data release 2, 2004). We
carefully checked our \hi channel maps 
for an \hi bridge, but even in the naturally weighted maps no connection
is visible. 
Figure~\ref{n4194chap_n4194hispec} shows the integrated \hi spectra of
NGC\,4194 
and the nearby galaxy derived from the data cube.  With an \hi line width of 
$\rm \sim 60 km\ s^{-1}$ and an \hi mass of $rm 7 \cdot 10^7\,M_{\odot}$ this
dwarf galaxy is significantly smaller than NGC\,4194 and thus does not play
an important role for the evolution of NGC\,4194.
%------------------------------------------------------------------------------------
%N4194 V band with beginning of HI tail FIG 6
%------------------------------------------------------------------------------------

%------------------------------------------------------------------------------------
%N4194 absorption FIG 7 
%-------------------------------------------------------------------------------------------------
%-------------------

\section{N-body simulations}

We have performed numerical simulations in order to test whether
the current appearance of the Medusa could be due to a gravitational
encouter between a gas-rich disk-like galaxy and an elliptical.
Our goal was not to reproduce every detail of the Medusa, but rather
to see whether 
a simple type of encounter could generate the observed overall distribution of the
stars and the gas, and in particular the following striking patterns:        
i) the diffuse region of stellar emission to the north (the ``hair");
ii) the tangential arcs seen in the stellar distribution to the south ;  
iii) the ca 50 kpc long tail of atomic gas to the south.

\subsection{The model}

We have used numerical simulations to follow the merging process of
two galaxies, a gas-rich disk-like galaxy (of spiral type) falling into
a larger spherical system (an ``elliptical" galaxy). 

The spherical galaxy is modeled by a set of collisionless particles 
%representing the stars 
distributed
according to the Hernquist profile (\citeyear{1990ApJ...356..359H}), % Hernquist 1990
with a  potential of the form $\Phi(r)= -GM/(r+a)$, 
where $G$ is the gravitational constant, 
$M$ is the mass, 
$r$ is the radius and $a$ is a scale length. 
That model has a simple density profile ($\rho\propto r^{-1}(r+a)^{-3}$) and produces a surface density 
profile in good agreement with the de Vaucouleurs $R^{1/4}$ law observed in 
elliptical galaxies, where $R$ is the projected radius on the sky. 
Particles are assigned random velocities calculated from the energy distribution function 
given by \cite{1990ApJ...356..359H}. % Hernquist 1990 

 For simplicity, the elliptical galaxy is represented by a single component 
(collisionless particles in a single Hernquist model)
and no distinction is made between stars and dark matter. 
Observationally, it is difficult to trace the mass distribution in elliptical 
galaxies out to large radii and although the amount of data is rapidly increasing 
(e.g., \citealt{2005MNRAS.357..691N}) % Napolitano et al. 2005
the scatter is large and the fraction of dark matter in ellipticals is still subject 
to interpretation
(e.g., \citealt{2003Sci...301.1696R}, %Romanowsky et al. 2003,
\citealt{2005Natur.437..707D}). %Dekel et al. 2005
No measurements of the stellar kinematics that would constrain on the mass-to-light 
ratio exist for the Medusa. 
In our model, the total mass of the elliptical is four times that of the infalling 
spiral. It corresponds to $6.66\times 10^{11} M_\odot$,
and about 50\% of that mass is contained within a radius of 12~kpc. In the following 
we call the collisionless particles ``stars", but it should be kept 
in mind that some of them, especially at large radii, could be regarded as dark matter 
particles. Since both the stars and dark matter particles 
have the same dynamics, having a mix of stars and dark matter woudn't affect the orbit 
of the infallling companion or the evolution of the system.

The companion galaxy consists of a stellar disk, a gaseous disk, and a dark matter
halo. The particles in the disk are distributed in two  
potentials of the form 
$\Phi(R,z) = -GM/\sqrt{R^2 + (a+\sqrt{z^2+b^2})^2}$, 
where $M$ is the mass, 
$R$ is the radius in the plane of the disk, 
$z$ is the coordinate perpendicular to the plane, and 
$a$ and $b$ are scale lengths 
\citep{1975PASJ...27..533M}. 
% Miyamoto and Nagai 1975
The advantage of that potential is that the scale lengths can be adjusted 
to represent different types of disks 
with a more or less prominent central bulge.  
A Hernquist model is used to represent the collisionless halo.  
Initially, the particles are assigned circular velocities
in centrifugal equilibrium with the gravitational potential,
and velocity dispersions are given by Toomre's stability criterion
\citep{1964ApJ...139.1217T}. % Toomre 1964

Initial conditions were generated from the distributions described above using Matlab. 
The evolution of the systems was computed by the programme {\tt COCKTAIL} 
(Collisions of Clouds Keep Tree Algorithm in Labor). 
The collisionless part is the tree algorithm of  
\cite{1986Natur.324..446B}. % Barnes and Hut 1986
The gas is modeled as an ensemble of clouds, all of 
the same mass, that dissipate energy through inelastic collisions
with each other (``sticky particles", e.g., 
\citealt{1985A&A...150..327C}, % Combes and Gerin 1985
\citealt{1997ApJ...481..132K}). % Kojima Noguchi
The gas accounts for
about 10\% of the mass of the stellar disk in the infalling galaxy. 

We used a total of 120000 particles: 30000 particles in the elliptical galaxy and 
90000 particles in the spiral  
(40000 in the stellar disk, 
20000 in the gaseous disk, 
30000 in the dark matter halo). 

The parameters of the simulation presented here are listed in
Table~\ref{tabsimul}.
Setting the units of length and time to 1~kpc and 10$^7$ years,
with the gravitational constant $G$ set to 1, gives a unit of
mass of $2.22\times10^9 M_\odot$.
We considered encounters of galaxies of different mass ratios and 
finally selected a mass ratio of 4:1, where the more massive galaxy 
is the elliptical. 
The initial mass of the gas in the spiral galaxy was set to match the observations of the Medusa, about 
$\rm 4.4\times10^9 M_\odot$. 
 The mass of a gas particle is therefore $\rm 2.2\times 10^5\,M_{\odot}$, which is of the
order of that a giant molecular cloud. 
The mass of a star particle in the disk galaxy is $\rm 10^6\,M_{\odot}$, and it is 
about $22\times 10^6$ in the elliptical galaxy.  
Ideally, all particles of a given type should have a comparable mass to prevent
artificial $N$-body scattering \citep{1942psd..book.....C}. In practice, however, this requirement is difficult to meet because 
of the limitation on the total number of particles set by an affordable computing time. 

Being mainly interested in the kinematical evolution of the stars and the gas in  
the infalling galaxy (and not in the internal relaxation of the merger remnant, which  
would be more sensitive to this issue), 
we used a comparatively larger number of particles to represent the stellar and the
gaseous components of the infalling galaxy.  
\\

The gas was initially distributed in a Miyamoto-Nagai disk with a scale length 
twice as large as that of the stellar disk, and the gaseous disk was truncated 
at an outer radius of 10~kpc. 
The scale length of the elliptical galaxy was calculated from the measured extent
of the Medusa at the 25th magnitude/arcsec$^2$ isophote: 
$D_{25} \times d_{25} = 2\farcm3\times 1\farcm6' = 26 \times 18$ kpc at
our adopted distance (RC3). 
This gives a radius to the 25th isophote of about 11~kpc. 
The scale length of the Hernquist profile is
related to the effective radius by the relation 
$R_e = 1.8153 a$  \citep{1990ApJ...356..359H}. % Hernquist 1990
For elliptical galaxies, $R_e < R_{25} < 2 R_e$. 
We thus took a scale length of 5~kpc for the elliptical galaxy.  

The two galaxies were initially separated by 50~kpc, and we tested different initial velocities. 
We obtained a best match for a slightly prograde encounter. In the simulation presented here, 
the spiral 
has initially a small positive tangential velocity 
(20~km~s$^{-1}$). 

%----------------------------------------------------------------------------------
\begin{table}[t]
\begin{tabular}{ll}
\noalign{\smallskip}
\hline
\noalign{\smallskip}
Mass ratio              & 4:1 \\
Elliptical: mass and scale length & M$_{\rm ell}$ = 300, $a_{\rm ell} = 5$\\
initial position and velocity   & $x=+10, y= 0, z= 0$\\
                                & $vx=0, vy= 0, vz= 0$\\
\noalign{\smallskip}
Disk galaxy:            & $M_{\rm stellar\, disk}  = 18$, $a_{\rm stellar\, disk} = 1$\\
mass and scale length   & $b_{\rm stellar\, disk} = 0.1$\\
                        & $M_{\rm gaseous\, disk} = 2$, $a_{\rm gaseous\, disk} = 2$\\
                        & $b_{\rm gaseous\, disk} = 0.1$\\
                        & $M_{\rm halo}  = 55$, $a_{\rm halo} = 5$\\
initial position and velocity   & $x=-40, y= 0, z= 0$\\
                                & $vx=0, vy= +0.2, vz= 0$\\
\noalign{\smallskip}
\hline
\end{tabular}
\caption[]{Parameters of the simulations, in computing units.
The units can be translated in the following physical units:
1~kpc, $10^7$~yr, 100~\kms, $2.22\times10^9 M_\odot$.
 }
\label{tabsimul}
\end{table}
%----------------------------------------------------------------------------------

\subsection{Results}

%---------------------------------------------------------------------------------
\begin{figure*}
\includegraphics[width=17.5cm]{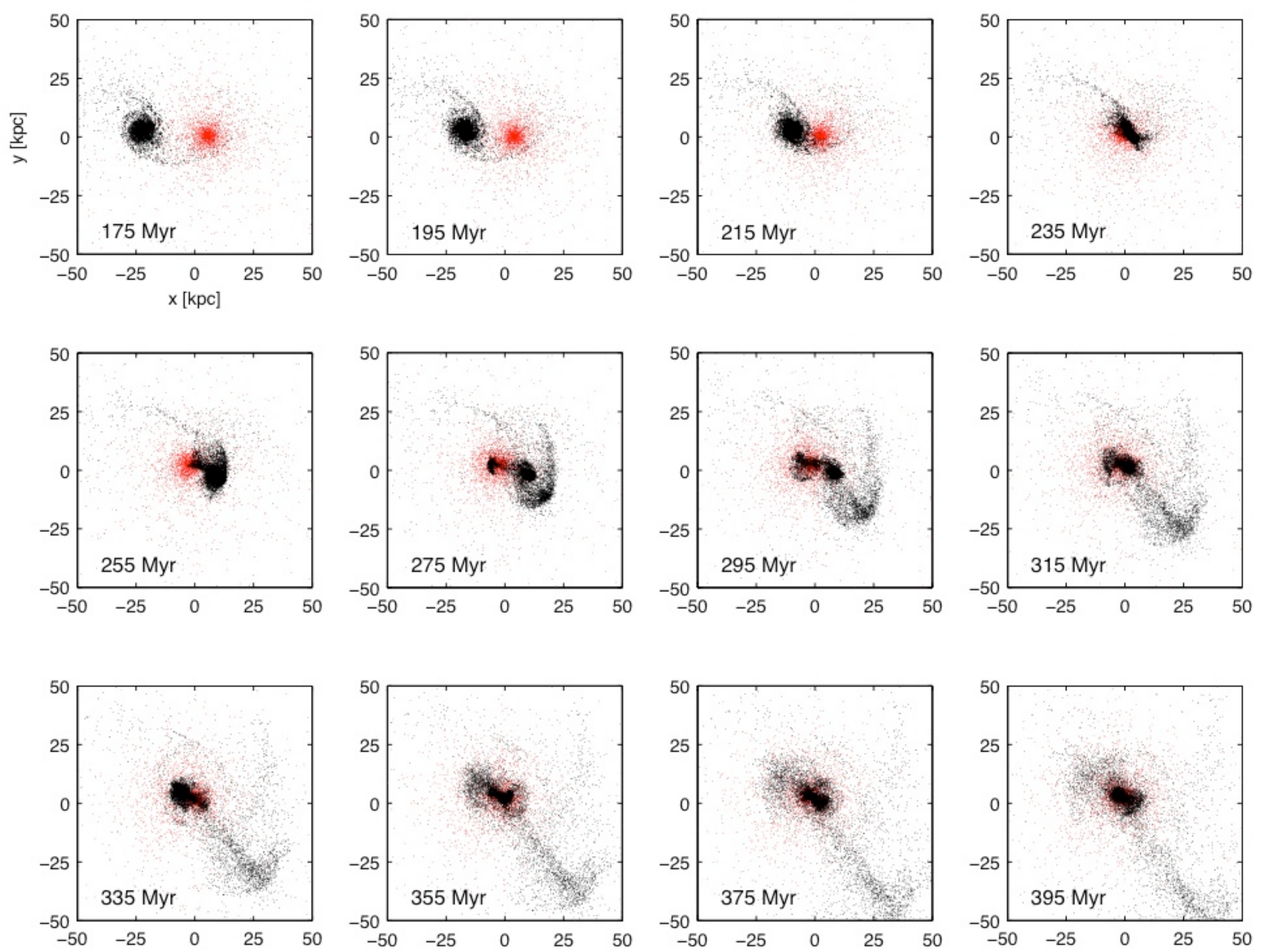}
\caption{Evolution of the stars in the simulation. The stars in the elliptical are shown in red,
those in the infalling disk galaxy in black. 
}
\label{figsimul1}
\end{figure*}
%---------------------------------------------------------------------------------
%---------------------------------------------------------------------------------
\begin{figure*}
\includegraphics[width=17.5cm]{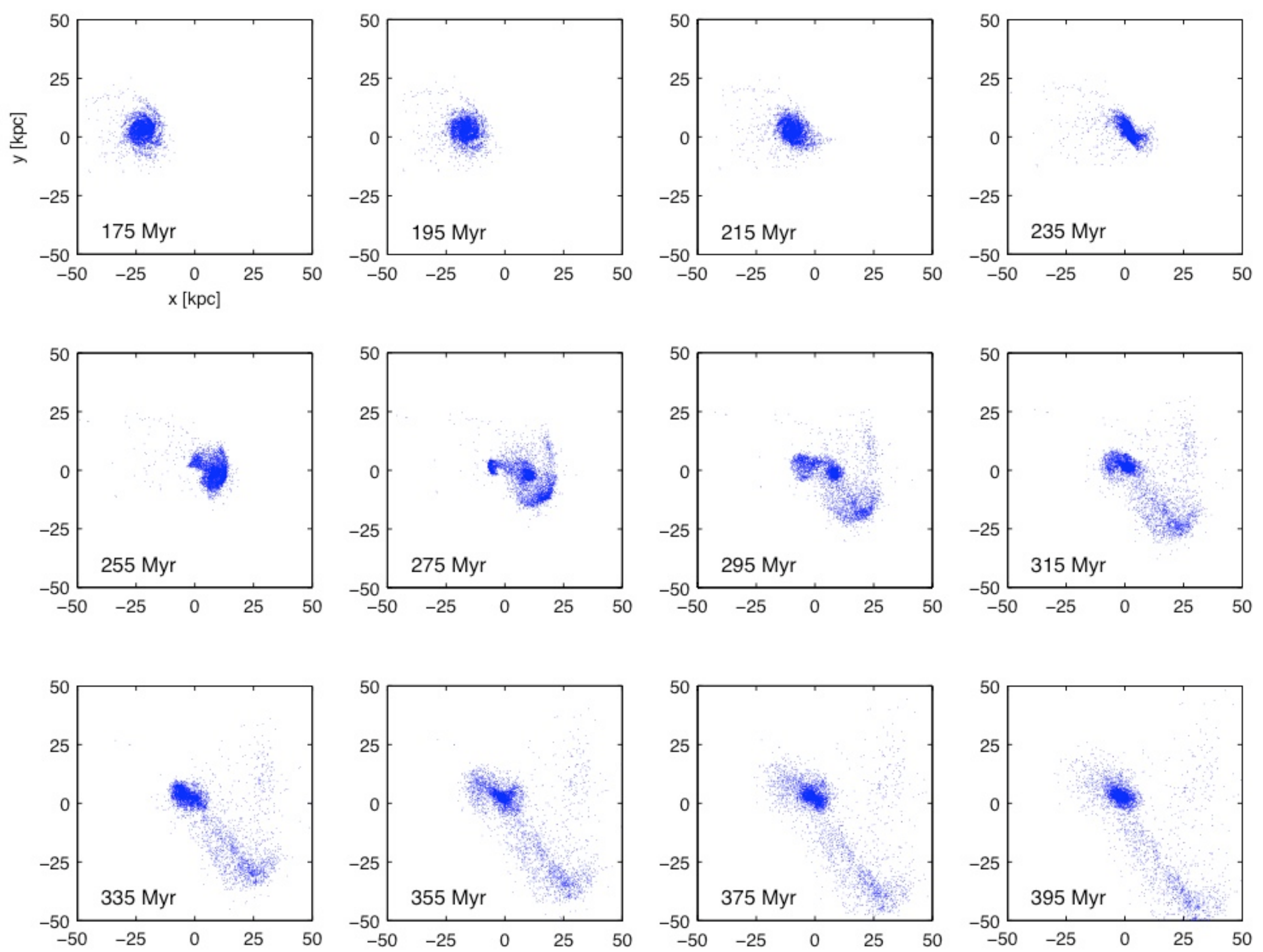}
\caption{Evolution of the gas in the simulation. 
}
\label{figsimul2}
\end{figure*}
%---------------------------------------------------------------------------------

Figures~\ref{figsimul1}  and \ref{figsimul2} display the evolution of the stars and the gas 
as the spiral galaxy falls into the larger elliptical. 
Tidal distortions appear
after closest passage and develop into a tidal tail, made of both stars and gas. 
The tail lengthens as the core of the infalling galaxy and the larger elliptical galaxy orbit around their center
of mass, leading to the eventual merging. Stars and gas are spread in the region diametrally opposite
to the tail, forming a diffuse distribution remiscent of the ``hair" of the Medusa. 
As particles from the spiral oscillate, shells form, due to the accumulation of particles at their apocenters.

The central part of the infalling disk galaxy is torn during the interpenetrating encouter with the 
more massive spherical galaxy. The material that will form the ``hair" crosses the center of the 
spherical galaxy twice and starts expanding into the diffuse ``hair" to the North while on the other side  
material expands into the tidal tail. 
The core of the infalling galaxy falls back through the center of the spherical a third time, and that is when 
the simulation resembles the observations most. Stars from the companion oscillate back and forth and 
shells and loops are seen, but they are transient and their evolution cannot be followed in detail because
of the relatively low number of particles that they contain. 

%-------------------------------------------------------------------------------
\begin{figure*}
\includegraphics[width=17.5cm]{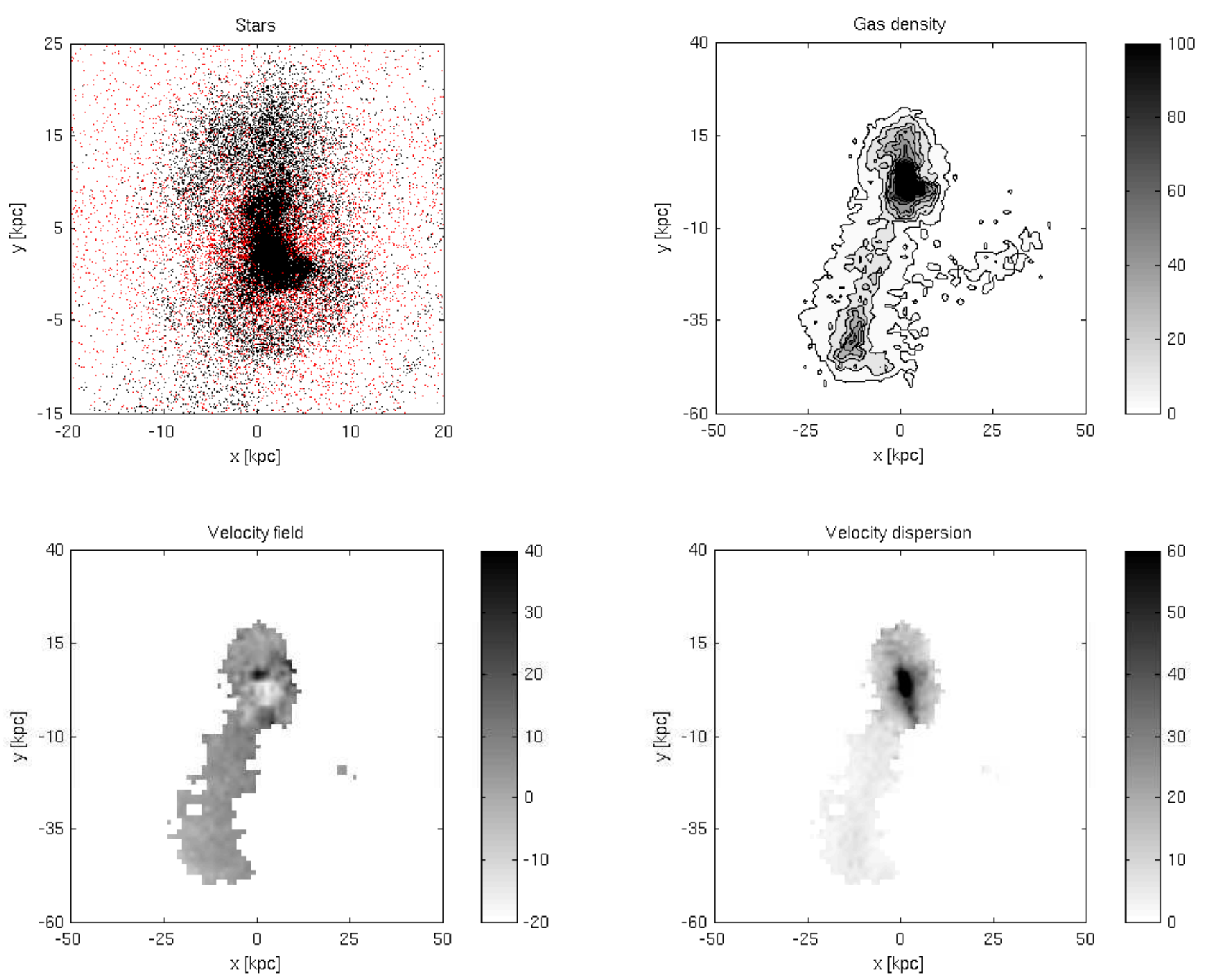}
\caption{View at a time corresponding to $T = 355$~Myr after the beginning
of the simulation, when the system resembles the Medusa most. 
The orbital plane (which is the same as the plane of the disk of the infalling galaxy)
also coincides with the plane of the sky. 
The system was rotated by 55$^\circ$ compared to the previous figures. 
a) Distribution of the stars. 
b) Distribution of the gas density. The grey scale shows the number of particles per 
1.5~kpc$^2$ pixel.
c) Velocity field of the gas. The grey scale shows the range of velocities in km~s$^{-1}$. 
d) Map of the velocity dispersion of the gas. The grey scale shows the range of velocity
dispersions in km~s$^{-1}$. 
}
\label{figsimul3}
\end{figure*}
%-------------------------------------------------------------------------------

Figure~\ref{figsimul3} shows the distribution of the stars and the gas and the kinematics of the gas 
at the time when 
the best match with the observations was found, $T = 355$~Myr after the beginning of the simulation, 
after a rotation of $+55^\circ$ in the plane of the galaxy. 
At that time, the length of the tail is about 50~kpc, and the kinematics are in reasonable agreement with the 
observations. 
Figure~\ref{figsimul3} can be compared both qualitatively and quantitatively 
to Fig.~\ref{n4194chap_n4194hi} that shows the observed counterparts. 
The observed \hi velocity field shows no gradient along the tail, as in 
the simulation. 
The velocity dispersion map is also shown. The highest values  
are found in the central body, whereas the tidal tail shows little
velocity dispersion.  
The isocontours show levels and constant particle density per pixel, where the pixel size of 
the simulated map is (1.5 kpc)$^2$. 
A density of 100 particles/pixel corresponds 
to $10^7 M_\odot$\ kpc$^{-2}$, or a column density of hydrogen of 
$1.25\times 10^{21}$~cm$^{-2}$, which translates into an \hi surface brightness 
of 0.5~Jy~km~s$^{-1}$\ beam$^{-1}$.
In the simulation, the levels of highest gas density are therefore about twice as high as the 
highest contour in the observed map. This was to be expected, since our total gas mass is exactly 
twice as large at the \hi mass and includes the molecular component, which is located in the central 
region of the galaxy. The simulation doesn't treat differently the atomic and the molecular gas. 
In the tail, the levels are comparable in the simulation and in the observed \hi map.
Defining the tail as the region with $y < -10$~kpc (see Fig.~\ref{figsimul3}), 
about 30\% of the gas particles and 20\% of the star particles lie in the tail. 
The gas fraction in the tail is comparable to the observed one. 
The amount of stars correspond to a surface brightness below the sensitivity of the
optical observations. 

\begin{figure*}
\includegraphics[width=8.75cm]{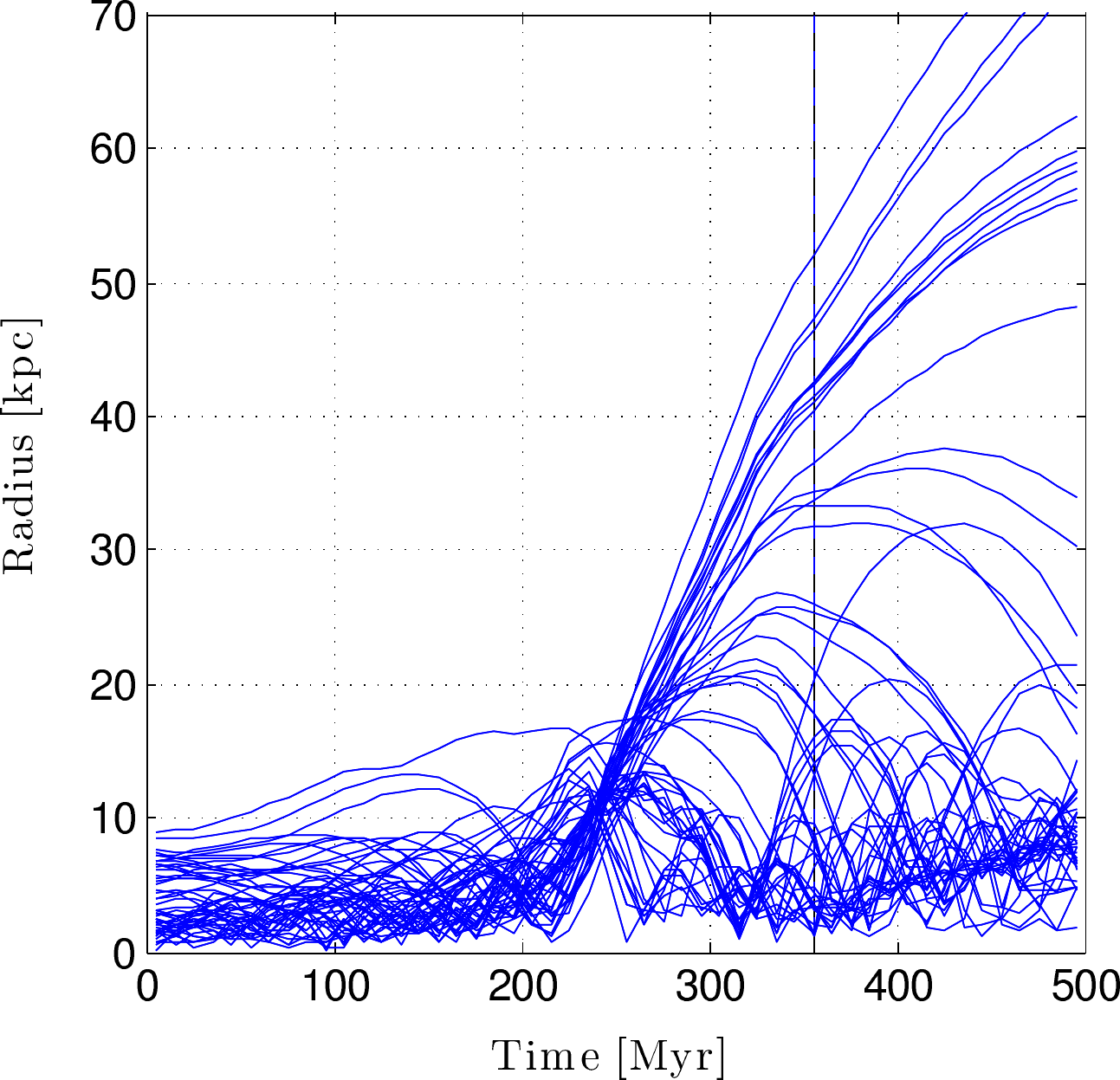}
\includegraphics[width=8.75cm]{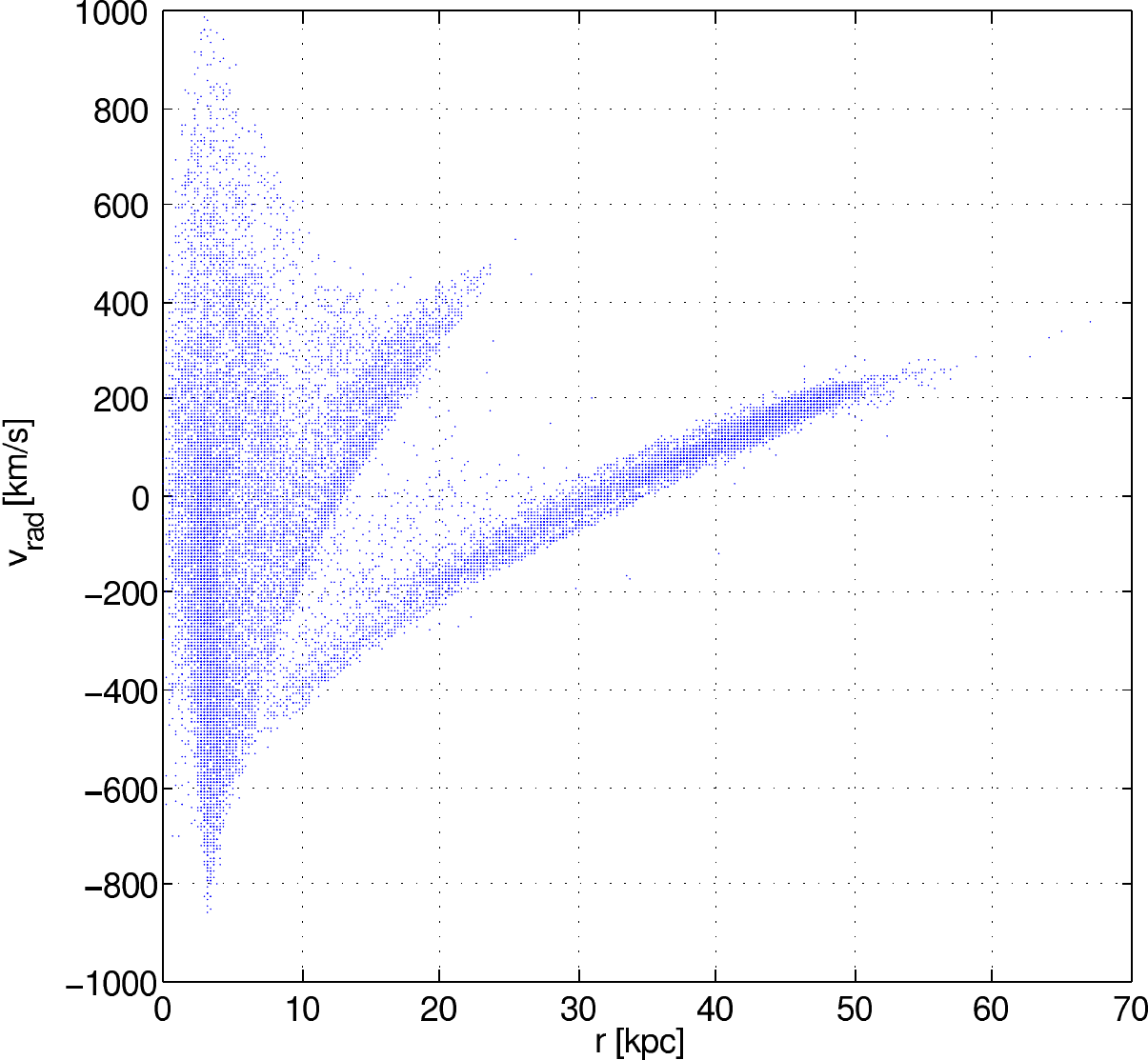}
\caption{
Kinematics of the gas in the simulation. 
Left: Temporal evolution of the radial distribution of the gas particles in the infalling galaxy. 
The vertical line at $T = 355$~Myr indicates the time when 
the system resembles the Medusa most. 
Right: Plot at $T = 355$~Myr of the radial velocity of the gas particles versus their distance from the center of mass of the infalling galaxy. 
The ``hair" and the tail appear as distinct kinematical features. 
}
\label{figsimul4}
\end{figure*}

Let us now examine the kinematics of the gas in the simulation. 
The particles forming the extended tidal tail to the South and those in 
the diffuse ``hair" to the North can be identified in the simulation by 
their different kinematics, as illustrated in 
Figure~\ref{figsimul4}. 
The left figure shows the radial oscillations of the gas particles in the 
infalling galaxy. The orbit crowding resulting in the tidal tail is clearly
visible, starting at $T = 210$~Myr. 
 The pattern is reminiscent to that due to radial waves in collisional ring galaxies 
(e.g. \citealt{1996FCPh...16..111A}, % Appleton \& Struck-Marcell 1996
\citealt{2001Ap&SS.276.1141H}). % Horellou \& Combes 2001
Here, however, the phenomenon is different because there is no quasi-perpendicular impact 
with respect to the plane of the disk galaxy that would cause an expanding ring wave. 
The crowded particles seen in Fig.~11  are not located at all azimuthal angles, but come  
from an unwinding arm that forms in the infalling galaxy  
after the crossing of the elliptical. 
At $T = 355$~Myr, when the system resembles
the Medusa most (the straight vertical line in the figure), most particles at 
a radius larger than 30~kpc are on their way out, wherease the ones at a smaller
radius are falling back toward the galaxy. This is seen in a different way in 
the figure to the right, which displays the radial velocity of the gas particles
versus their radius. This is related to the slope of the curves in the left figure. 
Particles can be separated into three groups: the tail, expanding beyond $r \simeq 30$~kpc 
($v_{rad} > 0$), 
the ``hair", expanding beyond $r\simeq 12$~kpc, 
and the rest, mostly located within $r \leq 10$~kpc. 

\begin{figure*}
\includegraphics[width=8.75cm]{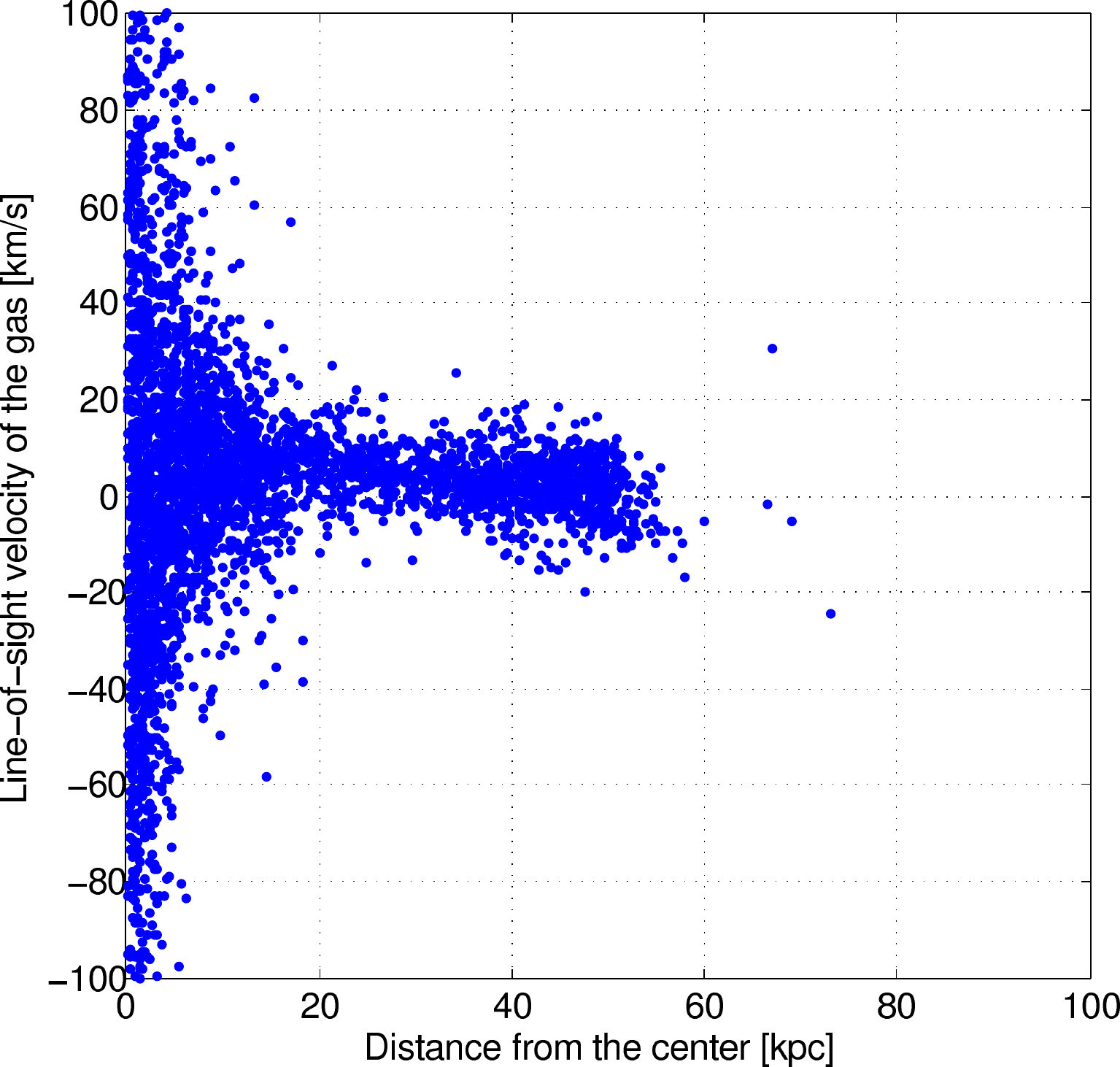}
\includegraphics[width=8.75cm]{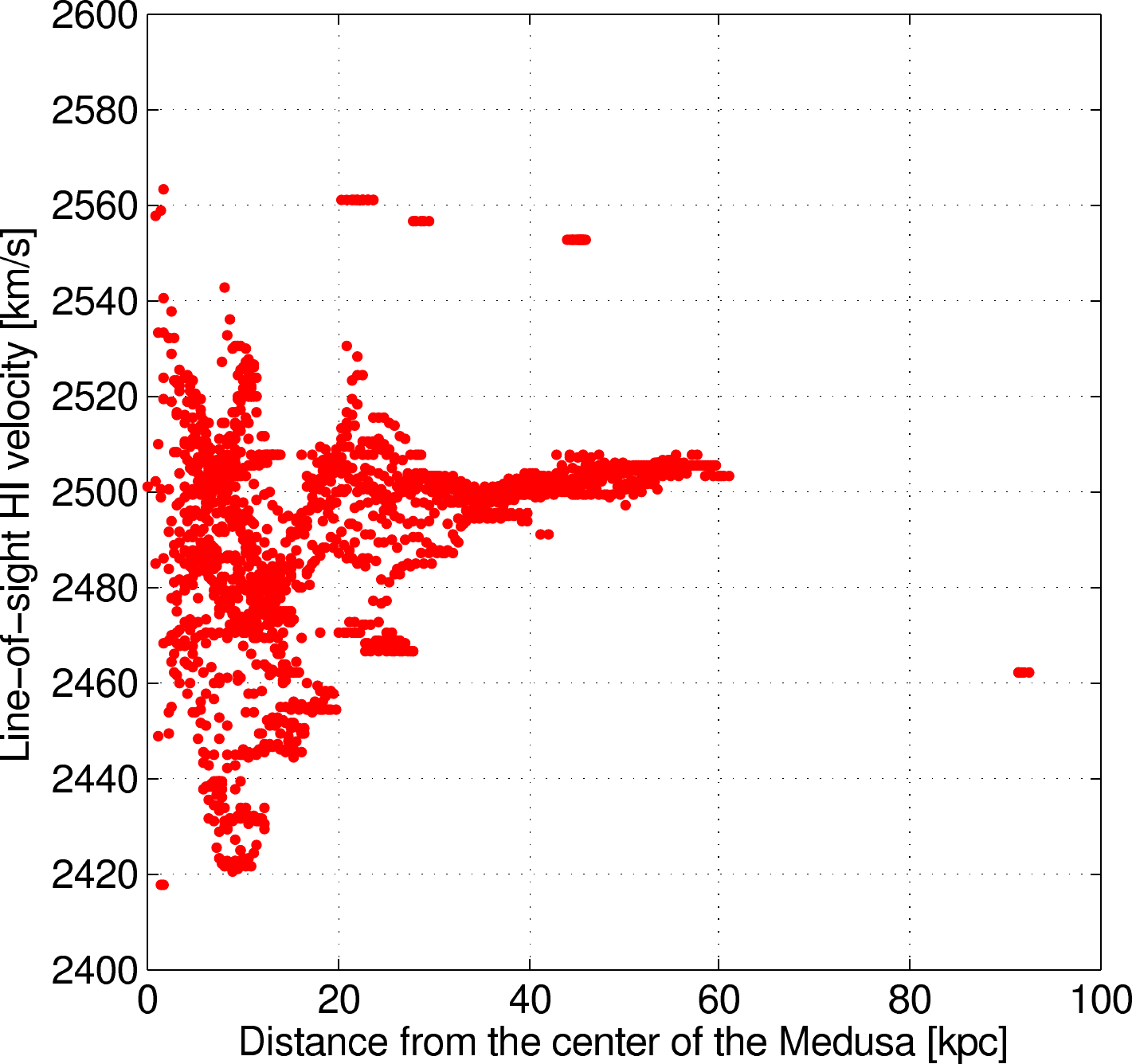}
\caption{
Line-of-sight velocity of the gas versus distance from a center in the simulation (left)
and in the observations (right). 
}
\label{figsimul5}
\end{figure*}

Since the different components have clear different kinematics, we would like to 
be able to identify them in the observed \hi data. 
Unfortunately, since the system is seen very close to face-on, the kinematical 
signatures which appear in the radial plots are not present when we examine the
line-of-sight velocity components, which are displayed in Figure~\ref{figsimul5} 
for both the simulation at $T = 355$~Myr (left) and the observations. 
The two plots are similar, displaying the gas tail extending to about 60~kpc with
a small and constant velocity dispersion as a function of radius. 
The simulation shows more gas in the center than the observation which is sensitive
to the atomic gas only, whereas the initial amount of gas in the simulation was 
taken to match both the molecular component (located in the inner part of the galaxy
and traced by CO observations) and the atomic component. 
%-------------------------------------------------------------------------------

%-------------------
\section{Discussion}

\subsection{Star Formation Rate}
We followed \cite{1992ARA&A..30..575C} and
\cite{2000ApJ...544..641H} and used the relation below to calculate the SFR
from the 20\,cm continuum flux:
\begin{center}
SFR ($M_{\odot}$\,yr$^{-1}$) = $0.14~D^2~F_{\rm 20\,cm}$ 
\end{center}
 where
$D$ is the distance in Mpc and $F_{\rm 20\,cm}$ is the
20-cm radio continuum flux density in Jy. The results are given in Table\,\ref{n4194chap_contflux}.
We derived a SFR which is comparable to the results of
\cite{1990ApJS...73..359C}, who measured a flux of 122\,mJy with the VLA and
thus 
determined the SFR in the Medusa as $\rm 
\sim 20\,M_{\odot}\ yr^{-1}$. 

%\begin{table}[h!]
\begin{table}
\caption[20\,cm continuum flux results of NGC\,4194]{The measured continuum
  flux, star formation rates based on continuum 
  and FIR emission, 
  respectively, and $q$ parameter.}            
\label{n4194chap_contflux}      
\centering                       
\begin{tabular}{l c }       
\hline\hline            
 Properties & \\   
\hline                  
 20\,cm (Jy) 			& 0.108\\
${\rm SFR_{20\,cm}}$ (\msun\ yr$^{-1}$)	& 22.8\\
${\rm SFR_{FIR}}$ (\msun\ yr$^{-1}$) 	& 9.5\\
$q$  				& 2.5\\
\hline                                   
\end{tabular}
\end{table}

Table~\ref{n4194chap_contflux} also gives the SFR estimated from the FIR luminosity:
\begin{center}
SFR ($M_{\odot}$\,yr$^{-1}$) = 0.17 $L_{\rm FIR}$
\end{center}
following \cite{1998ARA&A..36..189K}, with $L_{\rm FIR}$ in units of
$10^9$\Lsun. 
As a reference for the IRAS 60$\mu$m and 100$\mu$m flux densities in
Tab.\,\ref{n4194chap_intro} we used the
IRAS Faint Source Catalog (\cite{1990IRASF.C......0M}). 
The relation between the SFR estimated from the 20\,cm flux and the FIR fluxes
can be expressed by the $q$ parameter defined by
\cite{1985ApJ...298L...7H}, which is the logarithmic ratio of FIR to radio
flux. 
\cite{1985ApJ...298L...7H} derived a mean value of 2.3 for disk galaxies, 
independent of their
star formation activity.
Although NGC\,4194 is a merging galaxy, it conforms to this value which is a good
indicator that the continuum emission is dominated by emission from the starburst.
\cite{1985ApJ...298L...7H} mention that even galaxies with a strong nuclear
starburst, as Arp\,220, are in agreement with this correlation of the FIR and
radio flux, however
within a higher scatter. Arp\,220, e.g., has a $q$ value of 2.65,
whereas other prominent ULIRGs like Arp\,299 and Mrk\,231 have a value of
2.1. 

Our results are in good agreement with the investigations of
\cite{2004AJ....127.1360W}, who used the UV flux of the central star forming
knots to estimate the star formation rate. They estimate a SFR within the
knots of $\rm 5-6\,M_{\odot}\ yr^{-1}$ and suggest an overall SFR of up to $\rm
30\,M_{\odot}\ yr^{-1}$. 

The continuum source was studied recently by \cite{2005A&A...444..791B} with
 sub-arcsecond resolution MERLIN data. They found a pair of continuum
sources separated $\sim 0.35''$, thus unresolved in our image. They derived a
SFR of  $\rm 6\,M_{\odot}\ yr^{-1}$, in agreement with the 
results of \cite{2004AJ....127.1360W}, but they also
pointed out that they only measured the star formation rate in the very center
of NGC\,4194.

Our measured continuum source size of 1.2\,kpc is only slightly smaller than
the star forming 
region 
of $\sim$ 2\,kpc, as determined by optical and CO observations
(\citealt{1990ApJ...364..471A}, \citealt{2000A&A...362...42A}).

\subsection{The merger remnant}

Our simulations strongly support the hypothesis that the Medusa is 
the result of a merger between a gas-rich disk galaxy and a larger 
elliptical, with a mass ratio of about $4$. 
We have tested a wide range of parameters, and the constraints placed by 
the observations of the \hi tail (its length and kinematics) and the 
optical appearance of the Medusa made it possible to reduce considerably
the parameter space. 
The encounter that we finally selected is slightly prograde, in the sense that 
the orbital plane coincides with the plane of the infalling disk galaxy, but the
tangential velocity component is low -- 20\kms -- and the infalling galaxy acquires
largely radial motion as it falls through the larger elliptical. 
The two galaxies interpenetrate and the central part of the disk galaxy passes  
through the center of the elliptical twice, forming the diffuse ``hair" on one side. 
Our simulations are similar to those of 
\cite{1997ApJ...481..132K} % Kojima Noguchi
who investigated the formation of shells using a model 
similar to ours, where the gas is represented by ``sticky" particles. 
(Their model, however, included also star formation based on the local gas density 
and gas consumption). 
They showed that radial and slightly retrograde mergers lead to the formation of 
long-lived ($> 1$~Gyr) shells, and that star formation is significantly reduced 
because the gas is scattered in tidal debris. We have observed similar effects in some of the simulations
that we have run. We believe that the Medusa is due to a different type of encounter,
namely a slightly prograde one, also discussed by 
\cite{1997ApJ...481..132K},  % Kojima Noguchi
that leads to both the formation of an extended gas-rich tidal tail and to the 
infall of gas toward the center, where a moderate starburst is observed. 
Although the overall distribution of the stars and the gas in the simulation
is similar, some differences are seen. In low-density regions, the gas behaves like
the collisionless stellar component. During the interpenetrating encounter, the 
rate of collisions between the gas clouds increase, which lead to a loss of energy 
and angular momentum and an infall of gas toward the center; the gas and the stars 
have a different initial radial distribution, with the gas being located more in the outer
part with a larger angular momentum and lower binding energy. The gas 
is therefore more susceptible than the stars to be expelled into extended tidal tails. 

\cite{2000AJ....119.1130H} studied some examples of disk-disk mergers which
show anticorrelations between stellar and gaseous tidal features. They discuss
various possibilities for these differences, including ram pressure stripping,
photoionisation, dust obscuration and different origins in the progenitor
galaxies. Although possibly at least several scenarios may play a role at some
point, the lack of starlight associated with the large \hi tail found in
NGC\,4194 can be easiest explained by the original position of the \hi in the
progenitor. 

%------------------------------------------------------Conclusions---------
\section{Conclusions}  
We have presented new optical, radio continuum and \hi line observations of the Medusa galaxy and
numerical N-body simulations of the merger of a small 
gas-rich disk galaxy with a larger elliptical galaxy.  
The optical observations consist of a deep V-band image of the distribution of the stars in the Medusa, 
and the radio observations of maps of the 20 cm radio continuum emission 
and of the large-scale distribution and kinematics of the atomic hydrogen gas. 
The simulations are completely self-consistent with a total of 120000 particles. 
The gas in the infalling disk galaxy is represented by an ensemble of clouds, all of the same
mass, that can dissipate energy in inelastic collisions with each other.  

\begin{enumerate}
      \item The \hi observations revealed the presence of a single \hi tail to the South, extending 
      out to a projected distance of 4.95\,', or 56 kpc from the center of the Medusa.  The
      tail contains $\rm \sim 7\cdot 10^8\, M_{\odot}$ of \hi and has no 
      detected optical counterpart. 
      No significant velocity gradient is observed along the tail, 
      indicating that it is seen almost face-on.  

      \item \hi was detected in a dwarf galaxy identified at a projected distance
	of 91\,kpc to the North-West of NGC\,4194. Its systemic velocity of  
	2380\,km\ s$^{-1}$ is only 120\,km\ s$^{-1}$ lower than that of the Medusa. This dwarf
	galaxy harbours $\rm 
	\sim 7 \cdot 10^7\,M_{\odot}$ of \hi. No connection, neither in the
	optical nor 
	the \hi could be found between the dwarf galaxy and NGC\,4194 and it is 
        unlikely that it has had an influence on the dynamical evolution of the Medusa. 

      \item We found a large misalignment between optical and gaseous tidal
      features. The diffuse optical tail is situated on the opposite side of the  \hi  
      tail, the optical shells do not seem to be connected to the \hi 
      tail. Although the shells are within the region of the \hi  tail, no
      connection is apparent neither in morphology nor in velocity
      distribution. 

      \item We estimated a SFR of 
      23\,$\rm M_{\odot}\ yr^{-1}$
      based on the 20\,cm continuum flux density, and of 
      10\,$\rm M_{\odot}\ yr^{-1}$ from the FIR fluxes.  This is consistent with a 
      rather intense ongoing burst of star formation in the Medusa, 
      in agreement with the optical and UV observations of \cite{2004AJ....127.1360W} . 

      \item \hi was detected in absorption toward the radio continuum in the center.

      \item The simulations of a slightly prograde merger between a gas-rich 
disk galaxy with a total mass four times less than that of the elliptical 
give the best match to the observations. A extended single tidal tail develops 
in the outer part while the small galaxy falls in and oscillates around the 
center of the large elliptical, creating shell-like structures around the
elliptical, best visible in the stars. The overall distribution of the gas 
is similar to that of the stars, although more gas falls toward the center
because of the dissipative nature and more gas is found of the extended tail 
because of the initially more extended  distribution of the gas. 
      The tail is mostly formed by material of the outer \hi disk of
      the progenitor spiral galaxy, and the surface brightness of the stellar material 
in the tail is low. 

  \end{enumerate}

\begin{acknowledgements}
We thank the referee, Fr\'ed\'eric Bournaud, for a prompt and useful report.  
We thank the staff at the Westerbork Radio Synthesis Telescope 
for assistance during the observations,  
Tom Oosterloo and Gustaf Rydbeck for helping with some aspects of the data processing.  
We are grateful to Alessandro Romeo for useful discussions, to Joshua Barnes and to Piet Hut for making the treecode public and to 
Fran\c{c}oise Combes for providing us with a version of COCKTAIL. 
The Graduiertenkolleg 787 and the Swedish Research Council are acknowledged for funding support.  
This research has made use 
of NASA's Astrophysics Data System Bibliographic Services
and of the NASA/IPAC Extragalactic Database (NED) which is operated by the Jet Propulsion Laboratory, California Institute of Technology, under contract with the National Aeronautics and Space Administration.
\end{acknowledgements}

\bibliographystyle{aa}
\bibliography{0408}

\end{document}